\documentclass[twocolumn]{aa}

\usepackage{natbib}
\bibpunct{(}{)}{;}{a}{}{,}
\usepackage{graphicx}
\usepackage{txfonts}
\usepackage{lscape}
\usepackage{rotating}
\usepackage{longtable}
\usepackage{subfigure}

\begin{document}

\title{Stellar Population Gradients in Bulges along the Hubble
  Sequence.\thanks{Based
  on observations collected at the European Southern Observatory,
  proposals number 58.A-0192(A), 59.A-0774(A) and 61.A-0326(A) } }

\subtitle{II. Relations with Galaxy Properties}

\author{P. Jablonka \inst{1} \and J. Gorgas \inst{2} \and  P. Goudfrooij\inst{3} }

\institute{ Observatoire de l'Universit\'e de Gen\`eve, Laboratoire d'Astrophysique de
  l'Ecole Polytechnique F\'ed\'erale de Lausanne (EPFL), CH-1290 Sauverny,
  Switzerland
\and
Dpto. de Astrof\'{\i}sica, Facultad de F\'{\i}sicas,
 Universidad Complutense de Madrid, E-28040, Madrid, Spain
\and
Space Telescope Science Institute, 3700 San Martin Drive, Baltimore,
MD 21218, USA
}

\date{\today}

\abstract{
  
  We present   the  analysis  of  the   radial gradients  of   stellar
  absorption  lines   in  a sample of  32    bulges  of edge-on spiral
  galaxies,  spanning nearly the full Hubble   sequence (from S0 to Sc
  types),  and a large range of velocity dispersion (from about 60 to
  300 km    s$^{-1}$).  Different  diagnostics  such  as  index-index,
  gradient-gradient diagrams, and simple stellar population models are
  used  to tackle the origin   of the variation   of the bulge stellar
  population.   We find  that the vast  majority  of bulges show older
  age,  lower  metallicity  and higher   [$\alpha$/Fe]  in their outer
  regions than in their central parts.  The radial gradients in [Fe/H]
  are  2 to 3 times larger  than  in Log(age).  The relation
  between  gradient    and   bulge   velocity dispersion   is
  interpreted as  a gradual build up of  the gradient mean  values and
  their dispersions from high  to low velocity dispersion, rather than
  a pure  correlation.  The bulge effective  radii  and the Hubble
  type of  the  parent galaxies seem  to  play a more  minor  role in
  causing the observed spatial distributions.  At  a  given velocity
  dispersion, bulges and ellipticals share common properties.

}

\maketitle

\section{Introduction}

Combining results of several  recent studies of galaxies of  different
Hubble types, it seems clear that a proper  understanding of bulges of
spiral  galaxies is crucial   to  understanding  galaxy formation   in
general. It  has been realized  that early-type  (cD to E/S0) galaxies
harbor a variety of luminosity profiles, which has been interpreted as
being due to a varying contribution of a disk component (Saglia  et
al.\ \citeyear{1997ApJS..109...79S};  de Jong  et  al.
\citeyear{2004MNRAS.355.1155D}).     Bulges have  been 
envisaged as former ellipticals which reacted  to the presence of disk
accreted later on (e.g., Barnes \& White
\citeyear{1984MNRAS.211..753B}).  Other studies  have advocated  for a
``secular  evolution''  scenario to build bulges,  gradually inflating
them   using     disc     material      (e.g.,  Sheth      et     al.\ 
\citeyear{2005ApJ...632..217S} and references  therein).  For all such
studies,  one faces the  fact  that the  light distribution of (large)
galaxies is dominated by two components, bulge and disk (Allen et al. 
\citeyear{2006astro.ph..5699A}).  While  total disk magnitudes vary by
$\sim$\,2  mag  along  the  Hubble  sequence (taking   the faintest or
brightest  extremes of  the distribution at    each Hubble type),  the
bulge's ones   vary by twice  this  amount (Simien  \&  de Vaucouleurs
\citeyear{1986ApJ...302..564S};                   de              Jong
\citeyear{1996A&A...313...45D}), i.e., properties of bulges are key to
our understanding of the nature of the Hubble sequence.

With the exception of  the Milky Way and M31,  in which we can resolve
individual               stars        (Sarajedini     \&      Jablonka
\citeyear{2005AJ....130.1627S};    Olsen             et            al. 
\citeyear{2006AJ....132..271O}), studies of bulges  have to  deal with
integrated properties of  stellar populations. Despite the prospect of
yielding crucial information on  galaxy formation and assembly history
scenarios  from analyses of bulge    properties, bulges have  received
significantly less attention than E and S0 galaxies.  This is likely a
direct consequence of the considerable  challenge of avoiding possible
disk light contamination when studying bulges.

Spectroscopic studies of the metallicity and age  of the central parts
of  bulges were pioneered   by Bica (\citeyear {1988A&A...195...76B}). 
During the following several years,  clear evidence was provided for a
central metallicity-luminosity ($Z-L$)  relation for  bulges.  Several
studies  underlined  the similarities between  ellipticals  and bulges
(Jablonka,  Martin \& Arimoto \citeyear{1996AJ....112.1415J};  Idiart,
de Freitas Pacheco \& Costa \citeyear{1996AJ....112.2541I}), both with
respect to the slope of the $Z-L$ relation and to the similar apparent
$\alpha$-element overabundances  (relative  to solar abundance ratios)
in the two  types of  systems. The  Hubble  type of the parent  galaxy
appeared to have little influence on the stellar population properties
of  the bulge,   which  instead scaled   mainly with  central velocity
dispersion  and  bulge   luminosity  (Jablonka,   Martin   \&  Arimoto
\citeyear{1996AJ....112.1415J}).

A few years later, possible  distinctions were suggested between bulge
properties  of  late-type spirals  vs.\  those  of early-type  spirals
(Falc{\'o}n-Barroso,         Peletier             \&          Balcells
\citeyear{2002MNRAS.335..741F};         Proctor           \&    Sansom
\citeyear{2002MNRAS.333..517P}).  As pointed  out by  Thomas \& Davies
(\citeyear{2006MNRAS.366..510T})  however, these apparent  differences
seem to  disappear  when  the appropriate  range  of  central velocity
dispersion is considered  (in particular at  low  values). It  is also
likely true  that the homogeneity of the  samples and their sizes have
an impact on the drawn conclusions.

An important  limitation to our understanding  of bulge  properties is
that virtually  all  previous  spectroscopic  studies of   bulges only
sampled   the  light  in   the central  regions    which can easily be
significantly   affected by  disk   light,  especially  for  late-type
spirals.

Substantial progress  would    be enabled   with   spatially  resolved
spectroscopy of  bulges,  so that  radial gradients of  their  stellar
population can   be measured.   Unfortunately, investigations of  such
radial  gradients   using large  surveys have  so   far only addressed
early-type galaxies  i.e.,  elliptical and  lenticular  galaxies, with
only very modest   and rare excursions  into  the case of   later type
galaxies  (e.g., Sansom, Peace \& Dodd \citeyear{1994MNRAS.271...39S};
Proctor, Sansom \& Reid  \citeyear{2000MNRAS.311...37P}; Ganda et  al. 
\citeyear{2006MNRAS.367...46G};              Moorthy   \&     Holtzman
\citeyear{2006MNRAS.371..583M}).

In order to assess the extent of  the similarities between ellipticals
and   bulges of spiral  galaxies and  to evaluate  to what extent disk
material may be influencing the formation and  evolution of bulges, we
have undertaken a very deep spectroscopic survey of bulges in a large,
homogeneous  sample of  spiral  galaxies spanning most  of  the Hubble
sequence.

The galaxy sample,  the data reduction procedures  and the indices are
presented in detail in   Gorgas, Jablonka \&  Goudfrooij (\citeyear[][
hereafter   Paper I]{paper1}).  We present  the   analysis of the line
index radial changes  in the current paper.   It allows one at last to
establish (or discard) the   existence of scaling relations  and their
intrinsic dispersion, and to   shed some light   on the origin of  the
spatial variation of the stellar populations within bulges.

\section{The sample}

\begin{figure*}
\resizebox{1.\hsize}{!}{\includegraphics{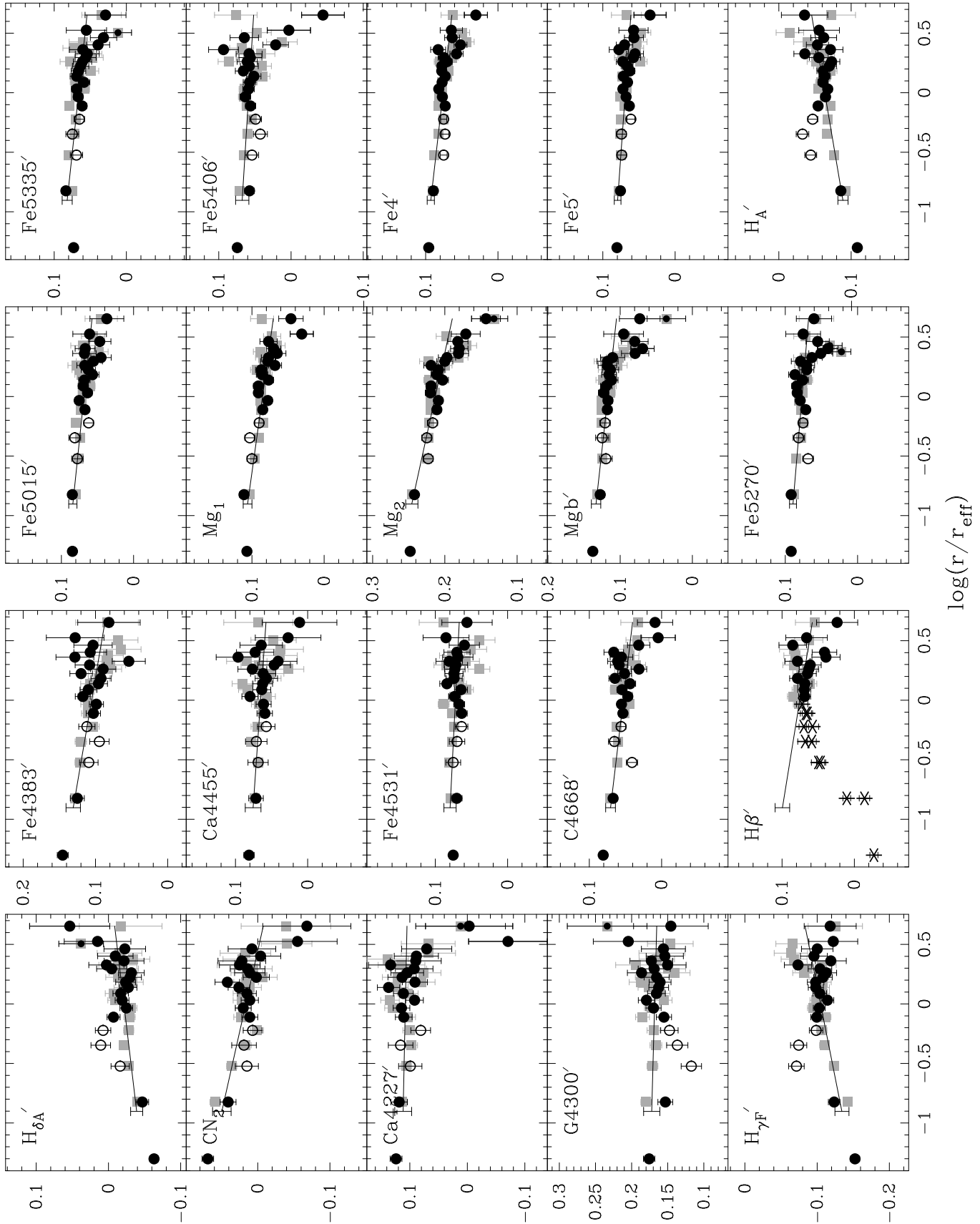}} \caption{Radial
variation of a sample of 20 spectral indices  for NGC 3957. Both sides
of the galaxy  minor axis are represented by   gray squares and  black
circles  for positive and negative  radii, respectively. The dust lane
is  identified  with open  symbols.  Dubious  data are indicated  by a
star. Indices have  been measured at  the  resolution of the  Lick/IDS
system. }
\label{ngc3957}
\end{figure*}

\begin{figure*}
\resizebox{1.\hsize}{!}{\includegraphics{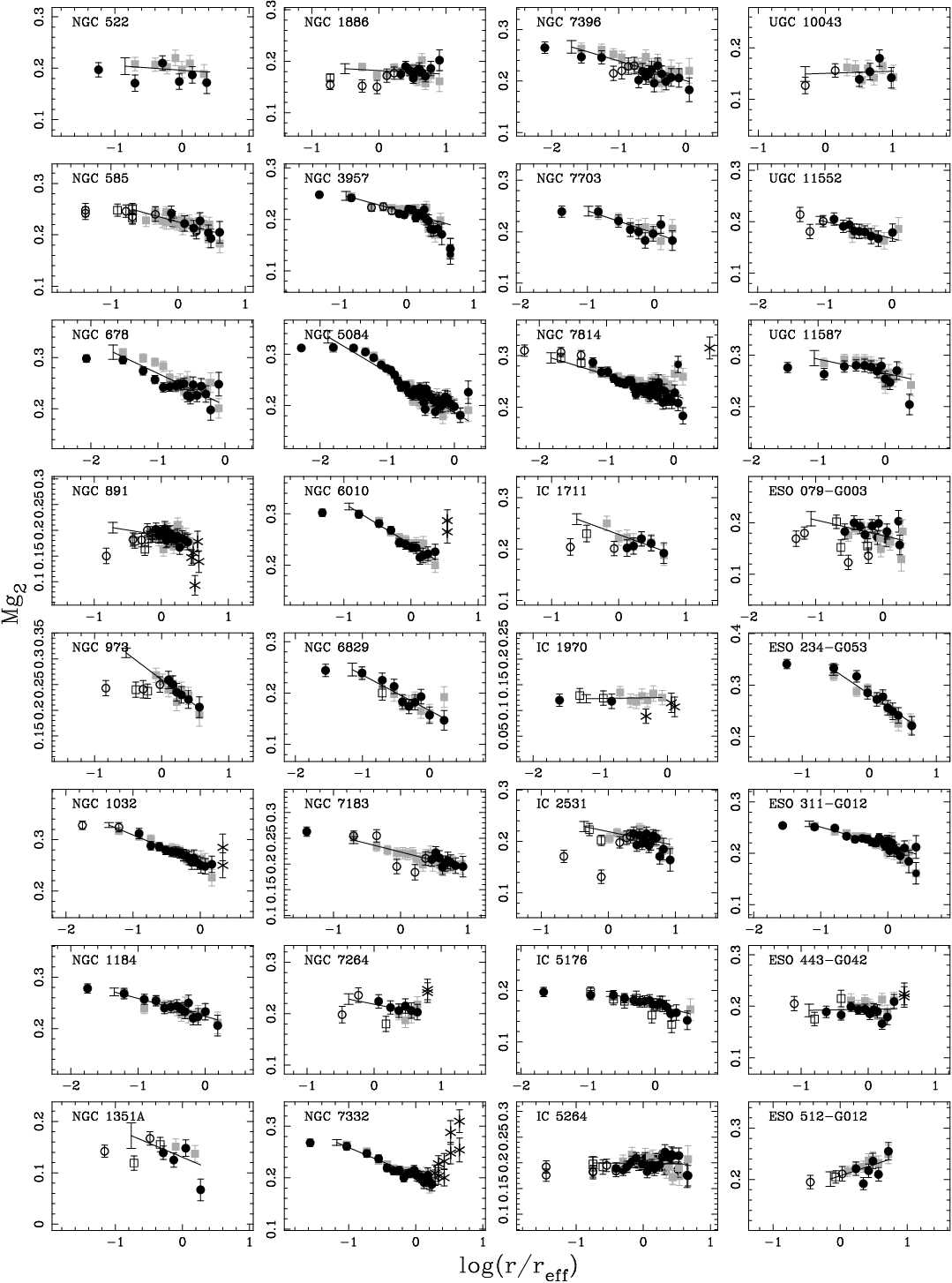}} \caption{The radial
change of Mg$_2$ for the full sample of galaxies. Symbols are the same as in
Figure 1. }
\label{example}
\end{figure*}

We selected a sample of 32 genuine (or close to) edge-on spiral
galaxies.  As shown in Paper I, 26 galaxies are inclined by 90
degrees, 6 galaxies have lower inclinations (60, 67, 75, 80 and 88
degrees).  Galaxies in the northern hemisphere were selected from the
Uppsala General Catalog (\citeyear{1973ugcg.book.....N}), while
southern galaxies were selected from the ESO/Uppsala catalog
(Lauberts
\citeyear{1982euse.book.....L}).  Hubble types  range  from S0 to  Sc,
with  the following frequencies: 7  S0, 4 S0/a, 2 Sa,  4  Sab, 9 Sb, 5
Sbc, and 1 Sc. Their  radial velocities range  from 550 to 6200  km/s.
Due  to the high inclination of  the galaxies, a precise morphological
classification is  difficult.  Therefore, we  assign an uncertainty of
about   one  Hubble  type, which   is  a  fair  representation  of the
catalog-to-catalog variations for a given galaxy.

Part of the   spectroscopic observations were  conducted   at the 2.5m
Isaac  Newton Telescope (INT) of the  Isaac Newton Group of telescopes
on the island of  La Palma (Spain). For the  southern galaxies we used
the 3.6-m ESO telescope and 3.5-m New Technology Telescope (NTT), both
at  ESO La   Silla observatory  (Chile).    The spectrograph  slit was
oriented along the minor axis of the bulges.

The      recent   work            of     Moorthy    \&        Holtzman
(\citeyear{2006MNRAS.371..583M}) is   the closest to  the
present study.  The size of their sample is  similar (38 galaxies vs.\
32) and they  also took care to span   a large range of Hubble  types.
However, Moorthy \& Holtzman's sample is composed of galaxies with any
inclination  angle,  and only about ten  of  them are  really close to
edge-on. This leads Moorthy \&  Holtzman to investigate the transition
between bulge  and disk populations, whereas we totally focus on
the intrinsic bulge properties.  The depth of the observations is also
different: our spectra reach  bulge regions at galactocentric
distances $\ga$ 3 times larger in radius than theirs.

{\tiny
\begin{table*}[h!p!t!]
\centering
\begin{tabular}{lrrrrrrrrrrrrr} \hline
 NGC  585  & H$\delta_A^\prime$ & H$\delta_F^\prime$ & CN1                & CN2                &  Ca4227$^\prime$   & G-band             & H$\gamma_A^\prime$ & H$\gamma_F^\prime$ & Fe4383$^\prime$    & Ca4455$^\prime$    & Fe4531$^\prime$    \\
  & $-$0.0086 &  0.0083 & $-$0.0216 & $-$0.0165 & $-$0.0304 & $-$0.0488 &  0.0395 &  0.0292 & $-$0.0208 & $-$0.0001 & $-$0.0176 \\
  &  0.0245 &  0.0255 &  0.0197 &  0.0198 &  0.0340 &  0.0136 &  0.0135 &  0.0216 &  0.0170 &  0.0167 &  0.0132 \\

& & & & & & & & & & \\

           &  C4668$^\prime$     & H$\beta^\prime$    & Fe5015$^\prime$    & Mg$_1$  & Mg$_2$ & Mgb$^\prime$  & Fe5270$^\prime$   & Fe5335$^\prime$  & Fe5406$^\prime$  & Fe5709$^\prime$    & Fe5782$^\prime$    \\
           & $-$0.0215 & $-$0.0459 & $-$0.0205 & $-$0.0180 & $-$0.0360 & $-$0.0189 & $-$0.0132 & $-$0.0238 & $-$0.0131 & $-$0.0101 & $-$0.0087 \\
           &  0.0116 &  0.0331 &  0.0094 &  0.0065 &  0.0064 &  0.0071 &  0.0075 &  0.0078 &  0.0109 &  0.0062 &  0.0130 \\
& & & & & & & & & & \\
          &  Na5895$^\prime$    & TiO1               & TiO2               & $<$Fe$>$$^\prime$      & Fe3$^\prime$       & Fe4$^\prime$       & Fe5$^\prime$       & Fe6$^\prime$       & H$_A^\prime$       & H$_F^\prime$       & Balmer  \\ 
           & $-$0.1275 & $-$0.0013 &  0.0056 & $-$0.0186 & $-$0.0202 & $-$0.0184 & $-$0.0133 & $-$0.0116 &  0.0155 &  0.0188 & $-$0.0341 \\
           &  0.0194 &  0.0045 &  0.0053 &  0.0061 &  0.0078 &  0.0067 &  0.0072 &  0.0053 &  0.0162 &  0.0195 &  0.0308 \\
 
\hline
\end{tabular}
\caption{For each of our sample galaxy, the 33 measured gradients (first line) and their attached errors (second line) are presented. The full table  is accessible along  with the electronic version of this paper.}
\label{tb:GradientsV}
\end{table*}
}
\section{The line-strength indices}

Our analysis is based on  the measurements of line-strength indices in
the           Lick/IDS     system          (e.g.,     Worthey       et
al.     \citeyear{1994ApJS...95..107W};   Worthey     \&     Ottaviani
\citeyear{1997ApJS..111..377W}).    For    the    sake  of  a   robust
determination  of the radial  variations  of the stellar population by
overcoming the  noise in   some individual  indices at faint   surface
brightness level  (as in the bulge  outskirts), we introduce a few new
indices  that  are  linear  combinations  of  some  classical Lick/IDS
ones. They are listed below:

\noindent $<$Fe$>$ = Fe2 = (Fe5270+Fe5335)/2; the classical mean Fe index.

\noindent Fe3 =  (Fe4383+Fe5270+Fe5335)/3; first introduced by Kuntschner (2000).

\noindent Fe4 = (Fe4383$+$Fe5270$+$Fe5335$+$Fe5406)/4 ; a composition of
indices     dominated    by    Fe   lines   (see      e.g.     Worthey
\citeyear{1998PASP..110..888W} for the chemical content of the individual indices).

\noindent Fe5 = (Ca4455$+$Fe4531$+$Fe5015)/3; based on indices dominated by a
mixture  of   metals,  including  $\alpha$ elements.

\noindent Fe6 = (Ca4455$+$Fe4531$+$Fe5015$+$Fe5709$+$Fe5782)/3; an
extended "mixed" composite Fe index. It could only be measured in galaxies
observed with the 3.6m ESO telescope (see below). 

\noindent H$_A$ = (H$\delta_A$$+$H$\gamma_A$)/2  and H$_F$ =
(H$\delta_F$$+$H$\gamma_F$)/2 ; two composite H indices.

\noindent Balmer = (H$\delta_A$$+$H$\gamma_A$$+$H$\beta$)/3; a composite Balmer
index, only useful for regions free of emission lines, since H$\beta$ is
included. 

In order to be consistent in our fitting procedures,  we use the same
units for  all indices,  converting  the index  $I$ into  the magnitude
index $I^\prime$ by:

  $$I^\prime= -2.5\log\left(1 - \frac{1}{n} \sum_{k=1}^n  I_k/ \Delta l_k\right)$$

\noindent where $n$ is the number of indices that compose the total index
$I$ and $\Delta l_k$ is the width (in \AA)  of the central passband in
which the index  $ I_k$ is  measured. Gradients in atomic indices  are
more linear with  galactocentric  radius when working  with $I^\prime$
than  with $I$.  Another advantage  of this strategy is  that the values of
all the  indices   are  comparable  with one   another, since  we  are
normalizing by the bandpass width.

We end  up with a grand total   of 33 different indices.   Because the
observations were done  on different telescopes and spectrographs, the
wavelength  coverage of  the spectra   is  not totally uniform   among
galaxies.  In particular, indices redder than 5600 \AA\ (i.e., Fe5709,
Fe5782, Na5895, TiO$_1$, and TiO$_2$) could be measured only for the 7
galaxies that were  observed at the ESO  3.6m telescope.  We refer the
reader to Paper~I for a complete  description of the conversion of the
indices to the Lick/IDS system.

\begin{figure*}[h!p!]
\resizebox{1.\hsize}{!}{\includegraphics{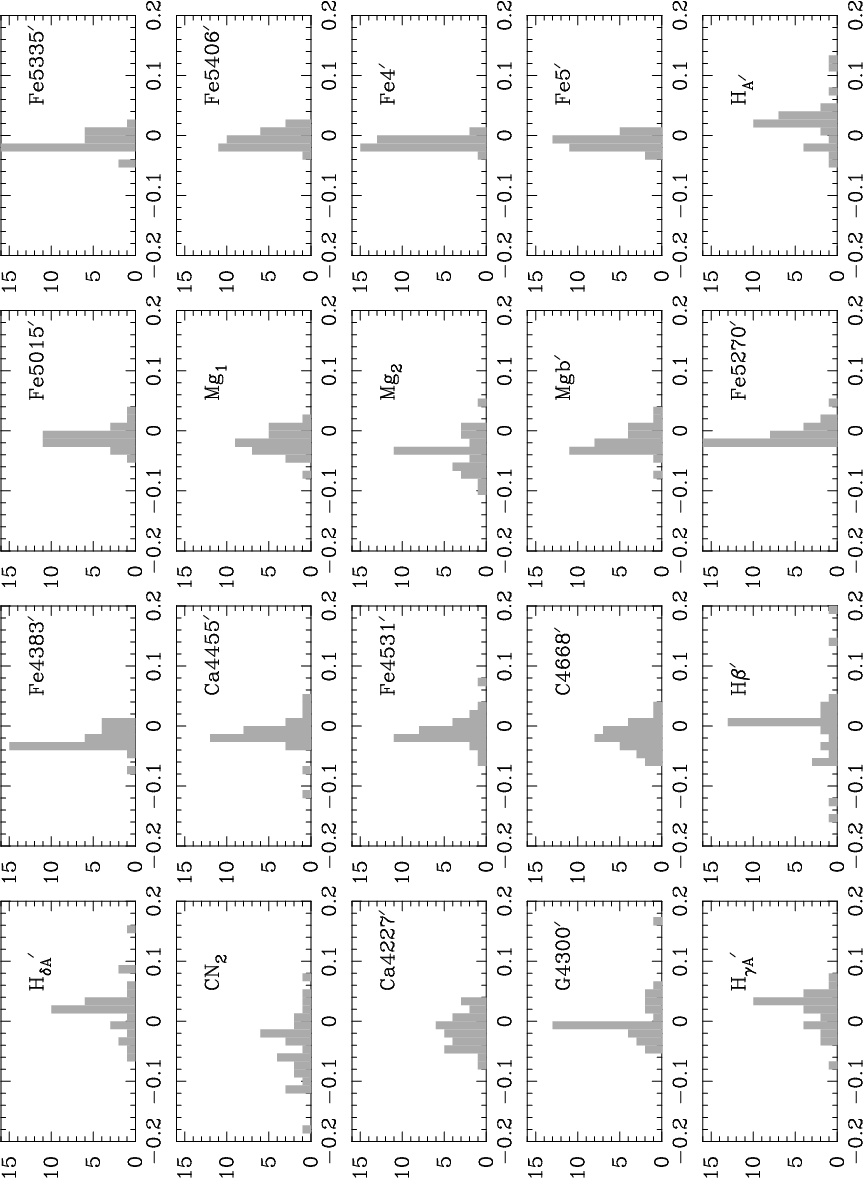}} 
\caption{Histograms (direct bining) of
the gradient amplitudes. The $x$-axis has been chosen such that the
extreme values of the gradients could be seen and that all indices could be
compared.}
\label{histo}
\end{figure*}

\section{The Gradients}
\subsection{Gradient Measurements}

For each galaxy,  the final radial sampling (i.e.,  the number of bins
along the  minor-axis radius) is  a  function of signal-to-noise ratio
(S/N).  Spatial rows  of the 2-D spectra were  summed until a  minimum
signal-to-noise of 10 per spatial bin was reached.  This procedure did
not concern the central parts,  where the S/N  values were much higher
even in a single spectrum row.  At the end of this procedure, a set of
1-D spectra  was  associated   with  each  galaxy, along  with   their
corresponding distances to the galaxy centre.

Bulge  minor  axis effective radii ($r_{eff}$)  were  derived  for all
bulges    as       explained  in    Paper~I.      As  illustrated   in
Figure~\ref{ngc3957} and Figure~\ref{example},  our spectra reach  the
bulge  effective radius for all our  sample galaxies.  For 25\% of the
sample, we reach 3~$r_{\rm eff}$ and for another 25\%, we could reach
10~$r_{\rm eff}$.     In order to  render  the  radial gradients fully
comparable from one galaxy   to another, we scaled all  galactocentric
distances  by these effective  radii, \[  I^\prime  = A  \log(r/r_{\rm
eff}) + B  \].  Radial gradients were derived  by  taking into account
both sides of  the  galaxy minor  axes,  avoiding dubious measurements
such as those  placed at the edges of  the CCD, excluding locations in
the  dust lanes, locations affected by  emission lines, as well as the
inner 0.5\,--\,1 arcsec (depending on the slit width) as they could be
affected   by disk  light   and seeing   effects.   The  limits of the
dust-obscured zone, translated in  a decrease in luminosity  along the
slits, have delineated the edges of the dust lanes.

We  adopted an  error-weighted least-squares  linear  fit to the data,
with a   two-iterations  procedure  to reject   points   falling beyond
$5\sigma$ from the mean relation. The gradients for our 33 indices and
32  galaxies  is accessible   with   the  electronic version   of  this
paper. Table~\ref{tb:GradientsV} provides an example of its format.

An example of the resulting gradients is presented in
Figure~\ref{ngc3957} for NGC~3957 and a subset of 20 indices.  The
plain lines indicate the fitted gradients and the error bar at the
left edge of the line shows the residual standard deviation of the
fit. Figure~\ref{example} displays the radial profile of the Mg$_2$
index for the full sample of galaxies and gives an idea of the variety
of gradients.  Galaxies like NGC 5084, which represents the
most extreme case, flatten in the inner regions.  For most galaxies,
however, a radial linear variation of the indices with radius is a
very fair representation.

\begin{table}[h!p!t!]
\centering
\begin{tabular}{lrccc} \hline
 Index               & Mean   & $\sigma_{rms}$  & $\sigma$ & $\sigma_{int}$ \\
       & (2)  & (3)           & (4)   & (5)            \\
\hline

H$\delta_A^\prime$    & 0.0236 & 0.0218 & 0.0123 & 0.0180         \\
H$\delta_F^\prime$    & 0.0159 & 0.0204 & 0.0149 & 0.0140         \\
CN$_1$                &$-$0.0618 & 0.0335 & 0.0126 & 0.0310       \\
CN$_2$                &$-$0.0635 & 0.0359 & 0.0154 & 0.0325       \\
Ca4227$^\prime$       &$-$0.0084 & 0.0148 & 0.0156 &              \\
G4300$^\prime$        &$-$0.0135 & 0.0183 & 0.0123 & 0.0135       \\
H$\gamma_A^\prime$    & 0.0339 & 0.0174 & 0.0095 & 0.0146         \\
H$\gamma_F^\prime$    & 0.0314 & 0.0226 & 0.0111 & 0.0197         \\
Fe4383$^\prime$       &$-$0.0265 & 0.0101 & 0.0102 &              \\
Ca4455$^\prime$       &$-$0.0162 & 0.0125 & 0.0106 & 0.0067       \\
Fe4531$^\prime$       &$-$0.0135 & 0.0108 & 0.0080 & 0.0072       \\
C4668$^\prime$        &$-$0.0295 & 0.0170 & 0.0072 & 0.0154       \\
H$\beta^\prime$       &$-$0.0039 & 0.0228 & 0.0190 & 0.0125       \\
Fe5015$^\prime$       &$-$0.0158 & 0.0083 & 0.0064 & 0.0053       \\
Mg$_1$                &$-$0.0264 & 0.0157 & 0.0058 & 0.0146       \\
Mg$_2$                &$-$0.0459 & 0.0205 & 0.0058 & 0.0197       \\
Mgb$^\prime$          &$-$0.0223 & 0.0127 & 0.0071 & 0.0105       \\
Fe5270$^\prime$       &$-$0.0149 & 0.0095 & 0.0066 & 0.0068       \\
Fe5335$^\prime$       &$-$0.0166 & 0.0087 & 0.0067 & 0.0055       \\
Fe5406$^\prime$       &$-$0.0145 & 0.0097 & 0.0069 & 0.0068       \\
Fe5709$^\prime$       &$-$0.0070 & 0.0064 & 0.0080 &              \\
Fe5782$^\prime$       &$-$0.0177 & 0.0112 & 0.0107 & 0.0034       \\
Na5895$^\prime$       &$-$0.0881 & 0.0632 & 0.0146 & 0.0615       \\
TiO$_1$               &$-$0.0025 & 0.0094 & 0.0060 & 0.0072       \\
TiO$_2$               &$-$0.0064 & 0.0117 & 0.0062 & 0.0100       \\
$<$Fe$>^\prime$       &$-$0.0163 & 0.0069 & 0.0051 & 0.0047       \\
 Fe3$^\prime$         &$-$0.0198 & 0.0075 & 0.0052 & 0.0054       \\
 Fe4$^\prime$         &$-$0.0186 & 0.0070 & 0.0043 & 0.0056       \\
 Fe5$^\prime$         &$-$0.0146 & 0.0076 & 0.0052 & 0.0056       \\
 Fe6 $^\prime$        &$-$0.0099 & 0.0052 & 0.0069 &              \\
 H$_A^\prime $        & 0.0293 & 0.0177 & 0.0077 & 0.0160         \\
 H$_F^\prime $        & 0.0239 & 0.0212 & 0.0093 & 0.0191         \\
 Balmer               & 0.0202 & 0.0180 & 0.0157 & 0.0088         \\
\hline
\end{tabular}
\caption{Statistics of the gradients for our set of indices. Column (2): error-weighted mean values of gradients; (3): dispersion
around the mean; (4): expected dispersion from the errors;
(5): intrinsic dispersion.} 
\label{tb:MeanGrad}
\end{table}

\subsection{Gradient Amplitudes and Statistics}

Figure~\ref{histo} presents     the   distribution of    the  gradient
amplitudes for our  main indices. The  values of the atomic  gradients
are small relative to past studies due to the logarithmic scale we are
using.  However, as seen in Figure~\ref{ngc3957}, the indices can vary
by 30 to 50 \% from the central to the  outer bulge regions.  The vast
majority of the   galaxies exhibit negative  gradients in  all indices
(except for the ones   involving Balmer lines), i.e.,  the  absorption
features increase  in strength towards the centre  of the bulges.  The
range  of gradient amplitudes is  quite similar from  one index to the
other.  Another   interesting feature  of Figure~\ref{histo} is  that,
despite  a  non-negligible  dispersion, the  distributions  are nearly
always well peaked, suggesting that  some factor other than randomness
is  at the  origin of the  spatial  distribution of  the bulge stellar
population properties.

Table~\ref{tb:MeanGrad} gives the error-weighted mean gradients and
the dispersion of their distributions for our 33 indices.  A
comparison between the mean values (column 2) and their attached
errors (column 3) demonstrates that the negative values of the
gradients are significant.  The dispersions of gradients among
galaxies are also significant and not due to observational errors as
indicated by the intrinsic dispersions. This means that there is a
variety of stellar populations in bulges and an intrinsic diversity in
their spatial variation, the origin of which we will try to unveil.

The dispersion  around  the   mean gradient  amplitudes  ($\sigma_{\rm
rms}$)  decreases with  redder indices (i.e.,   those measured in  the
redder part of  the bulge spectrum).  While  this is partly due to the
measurement errors which get smaller at larger wavelength, this cannot
fully explain the  measured dispersions, as  can be  inferred from the
intrinsic ones.   The indices  which  measure a given class of
chemical elements (i.e., $\alpha$ elements, iron, CNO) present similar
intrinsic  dispersions.    At  this  stage,   it is difficult   to
identify the origin of the differences in intrinsic dispersion between
the different indices.  They can  have two and possibly mixed origins:
varying ratios of the different  chemical species, but also, at  least
partly,  difference   in the dynamics  of    the indices, i.e.   their
relative sensitivity to changes in abundance or age.

\section{The Scaling relations}

\begin{figure*}[h!p!]
\resizebox{1.\hsize}{!}{\includegraphics{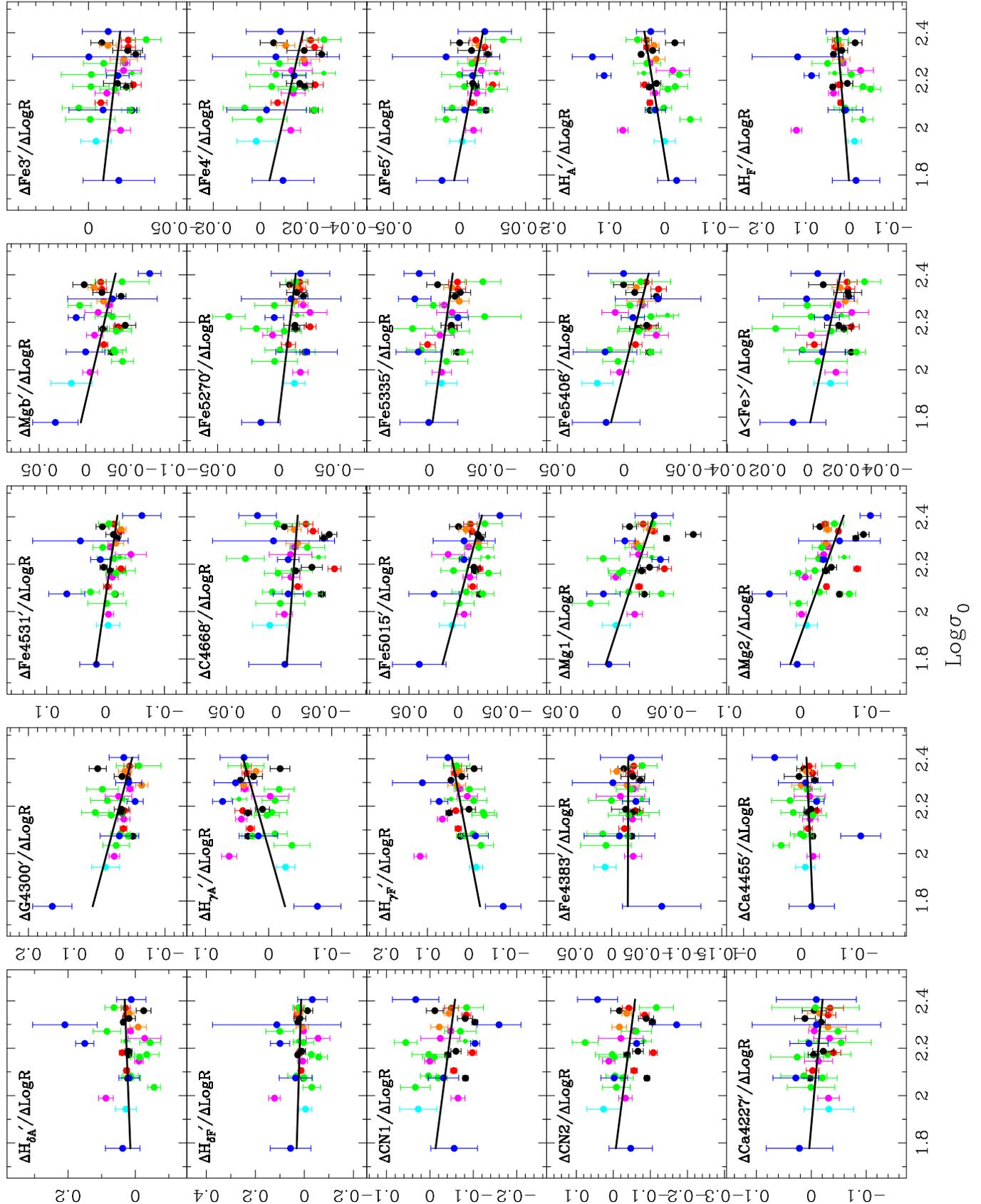}}
\caption{The relation between the gradients and the bulge central
  velocity dispersions for the 25  indices  that are available for  the
  whole  sample of  bulges. The  color  code for  the different Hubble
  types is  the following:  Black corresponds to   E/SO, red to  S0/a,
  orange to Sa, magenta to Sab, green to Sb,  blue to Sbc, and cyan to
  Sc types. The black solid lines show the linear fits.}
\label{sigma}
\end{figure*}

\begin{figure*}[t!p!h!]
\centerline{\includegraphics[width=1.0\textwidth]{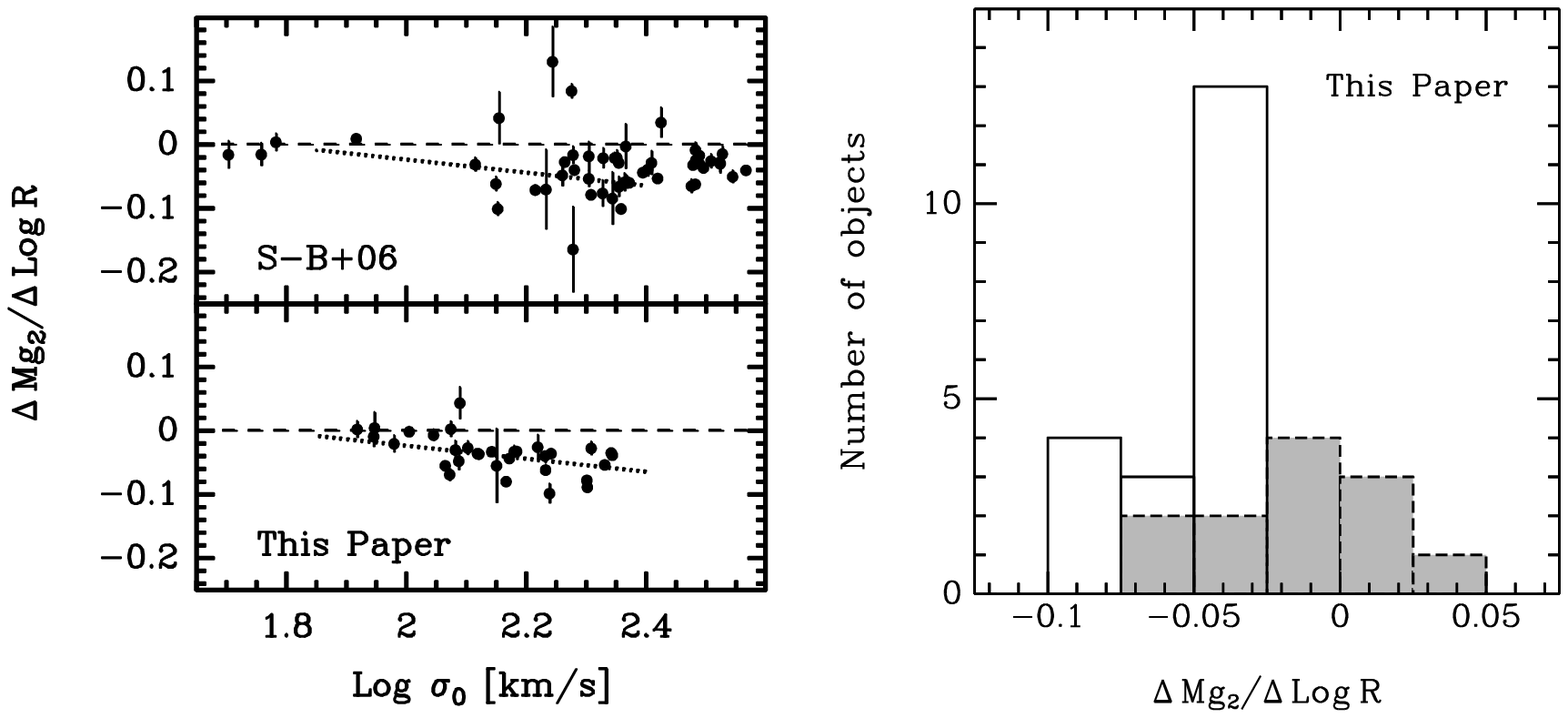}}
\caption{{\it Left panels}: The relation between Mg$_2$ gradient amplitudes and
central  velocity dispersion for bulges    in this paper (lower   left
panel) and  for       elliptical    galaxies in the     sample      of
S\'anchez-Bl\'azquez et al.\ (2006;   upper left panel). To guide  the
eye, a least-squares fit  to the data for  the bulges in this paper is
plotted  as  dotted lines in   both  left panels.   {\it Right panel}:
Histograms  of  Mg$_2$  gradient  amplitudes  for  bulges with central
velocity dispersions $\sigma < 125$ km s$^{-1}$ (dashed lines and grey
hashing) and with $\sigma \geq 125$ km s$^{-1} $ (solid lines).}
\label{Mg2Sigmaplot}
\end{figure*} 

As seen in the previous section, bulges, in general, do exhibit radial
changes in their stellar population  properties: their central indices
are  the  strongest.  In   this section,   we investigate whether  the
strength of the gradients   scales with some structural parameters  or
dynamical properties of the bulges.

\subsection{The existence of the gradient--$\sigma_0$ relation}

Figure \ref{sigma} presents  the relation between the gradient
amplitudes and the bulge central velocity dispersion ($\sigma_0$, in 
km~s$^{-1}$), calculated in a 2$''$-radius aperture. The solid  
lines indicate the error-weighted  least-square  linear fits: 
 \[  \frac{\Delta  I^\prime}{\Delta\log(r/r_{\rm  eff})}  =  A
\log( \sigma_0) + B \].

Table \ref{tb:GradSig}  provides  the coefficients $A$  and $B$, along
with the  error-weighted   standard deviation $\sigma$.   In order  to
evaluate the significance of the  fits,  we performed an F-test  which
checks the validity of the assumption of  a description of the data by
a flat relation fixed  to the mean of the  gradients.  It compares the
variance  of the distribution implied  by  such an  assumption to  the
observed one.  The smaller the probability returned by the F-test, the
more different the distributions. We also  used the Spearman test,
which qualifies the significance  of the correlation between the  gradients
and $\sigma_0$.  This test does not take the observational errors into
account. This is why  we use two  independent statistical tests.  The
smaller the Spearman probability, the more significant the correlation
between  the gradients and  the bulge central velocity dispersions. As
also    shown  in Figure \ref{sigma},  there    seems to be  no
statistically significant variation of  the amplitude of  the stellar
population gradients with the  bulge central velocity dispersion, with
the exception  of the Mg$_1$ and  Mg$_2$ indices.

\begin{table}[p!h!t!]
\centering
\begin{tabular}{lrrccc} 
\hline  
 Index               &  $A$~~~~~~~~          & $B$~~~~~~~~         & $\sigma$  & F  & S      \\
\hline
H$\delta_A^\prime$   & 0.025$\pm$  0.021  &$-$0.032$\pm$  0.047  &  0.02  & 0.97 & 0.81    \\
H$\delta_F^\prime$   & 0.028$\pm$  0.025  &$-$0.045$\pm$  0.055  &  0.02  & 0.95 & 0.98    \\
CN1                  &$-$0.046$\pm$  0.022  & 0.039$\pm$  0.049  &  0.03  & 0.95 & 0.17    \\
CN2                  &$-$0.033$\pm$  0.026  & 0.010$\pm$  0.057  &  0.04  & 0.98 & 0.11    \\
Ca4227$^\prime$      &$-$0.058$\pm$  0.026  & 0.117$\pm$  0.056  &  0.02  & 0.56 & 0.20    \\    
G4300$^\prime$       &$-$0.036$\pm$  0.020  & 0.065$\pm$  0.044  &  0.02  & 0.89 & 0.02    \\  
H$\gamma_A^\prime$   & 0.042$\pm$  0.015  &$-$0.059$\pm$  0.034  &  0.02  & 0.83 & 0.10    \\     
H$\gamma_F^\prime$   & 0.042$\pm$  0.017  &$-$0.061$\pm$  0.038  &  0.03  & 0.89 & 0.05    \\   
Fe4383$^\prime$      &$-$0.015$\pm$  0.015  & 0.008$\pm$  0.034  &  0.01  & 0.92 & 0.37    \\ 
Ca4455$^\prime$      &$-$0.005$\pm$  0.017  &$-$0.006$\pm$0.036  &  0.01  & 1.00 & 0.65    \\  
Fe4531$^\prime$      &$-$0.021$\pm$  0.013  & 0.032$\pm$  0.028  &  0.01  & 0.89 & 0.06    \\ 
C4668$^\prime$   &$-$0.009$\pm$  0.012  &$-$0.009$\pm$0.026  &  0.02  & 0.99 & 0.53    \\ 
Fe5015$^\prime$      &$-$0.003$\pm$  0.011  &$-$0.009$\pm$ 0.025 &  0.01  & 1.00 & 0.08    \\
Mg1                  &$-$0.066$\pm$  0.010  & 0.121$\pm$  0.021  &  0.01  & 0.49 & 0.01    \\
Mg2                  &$-$0.053$\pm$  0.009  & 0.071$\pm$  0.020  &  0.02  & 0.79 & 0.00    \\
Mg$_b^\prime$        &$-$0.004$\pm$  0.012  &$-$0.014$\pm$ 0.026 &  0.01  & 1.00 & 0.42    \\ 
Fe5270$^\prime$      &$-$0.009$\pm$  0.010  & 0.006$\pm$  0.023  &  0.01  & 0.97 & 0.33    \\   
Fe5335$^\prime$      &$-$0.008$\pm$  0.011  & 0.002$\pm$  0.024  &  0.01  & 0.97 & 0.15    \\   
Fe5406$^\prime$      &$-$0.019$\pm$  0.010  & 0.027$\pm$  0.023  &  0.01  & 0.87 & 0.08    \\   
$<$Fe$>^\prime$      &$-$0.014$\pm$  0.008  & 0.015$\pm$  0.018  &  0.01  & 0.88 & 0.13    \\
 Fe3$^\prime$        &$-$0.014$\pm$  0.008  & 0.011$\pm$  0.018  &  0.01  & 0.90 & 0.15    \\
 Fe4$^\prime$        &$-$0.012$\pm$  0.007  & 0.008$\pm$  0.014  &  0.01  & 0.90 & 0.07    \\
 Fe5$^\prime$        &$-$0.007$\pm$  0.008  & 0.002$\pm$  0.018  &  0.01  & 0.97 & 0.06    \\
 H$_A^\prime$        & 0.026$\pm$  0.013  &$-$0.027$\pm$  0.028  &  0.02  & 0.94 & 0.07    \\
 H$_F^\prime$        & 0.036$\pm$  0.015  &$-$0.054$\pm$  0.033  &  0.02  & 0.92 & 0.16    \\
\hline 
\end{tabular}
\caption{Linear fits of the variation of the gradients with the bulge central velocity dispersion   for the full galaxy  sample. $A$   stands for  the slope  of the
  relations, $B$ for their zero points, and $\sigma$ is the dispersion of
  the fits.  F and S  give the probabilities  of the F- and Spearman ,
  respectively (see text).}
\label{tb:GradSig}
\end{table}

Intriguingly, while Mg$_1$ and Mg$_2$ (both of  which are sensitive to
magnesium abundance)  vary  with  $\sigma_0$,   Mgb does not.     This
dichotomy between  the behaviors of Mg$_2$  and  Mgb has  already been
noticed    for    elliptical     galaxies.         Mehlert et      al.
(\citeyear{2003A&A...407..423M})   concentrate  on  Mgb   gradients in
elliptical  galaxies with a range   of velocity dispersions similar to
ours and  do       not  find  any    correlation.   More    recently,
S\'anchez-Bl\'azquez et al.  (\citeyear{2006A&A...457..823S})  confirm
this   absence   of   correlation.    Meanwhile,  Carollo   et  al.
(\citeyear{1993MNRAS.265..553C}) reported on the correlation of Mg$_2$
gradients with $\sigma_0$  among  elliptical galaxies.  Ogando  et al.
(\citeyear{2005ApJ...632L..61O})  also observe a  relation between the
Mg$_2$ gradients and the velocity dispersions of their galaxies.

If one considers the simple stellar population (SSP) models of
Vazdekis (\citeyear{1999ApJ...513..224V}) or Thomas et al.
(\citeyear{2003MNRAS.339..897T}), the sensitivity of Mgb$^\prime$ to
age at fixed metallicity (or to metallicity at fixed age) is
$\sim$\,half that of Mg$_2$.  Indeed, our Mgb$^\prime$ gradients are
half as large as the Mg$_2$ ones on average.  This is however not
sufficient to explain the change in behavior of the two indices.  It
has to be attributed to the intrinsic properties of the indices and,
in particular, to their dependencies to abundances.  As a clear
illustration of this, four galaxies (UGC~10043, ESO~443-042,
ESO~512-012, and IC~1970) do not exhibit any gradient in Mg$_1$ nor in
Mg$_2$, but they do have significantly negative Mgb$^\prime$
gradients.  In those cases, C4668$^\prime$ shows no radial change either.
Therefore, for those galaxies, the carbon abundance seems to supersede
the magnesium sensitivity of the indices Mg$_2$ and Mg$_1$. This Mg
vs. C balance is seen in a quite spectacular way for these 4 galaxies,
but it could play a (likely more subtle) role in other bulges.  In any
case, the sensitivity to different chemical elements seems to be able
to explain the discrepant behavior of the three ``magnesium'' indices.

The  indices  primarily sensitive  to    iron exhibit gradients  whose
amplitudes seem independent of the bulge  velocity dispersion.  As can
be seen in Table \ref{tb:MeanGrad}, the intrinsic dispersions of these
amplitudes  are extremely  small and leave    hardly any room  for any
variation beyond the one consistent with the uncertainties.

\begin{figure*}[t!p!h!]
\resizebox{1.\hsize}{!}{\includegraphics{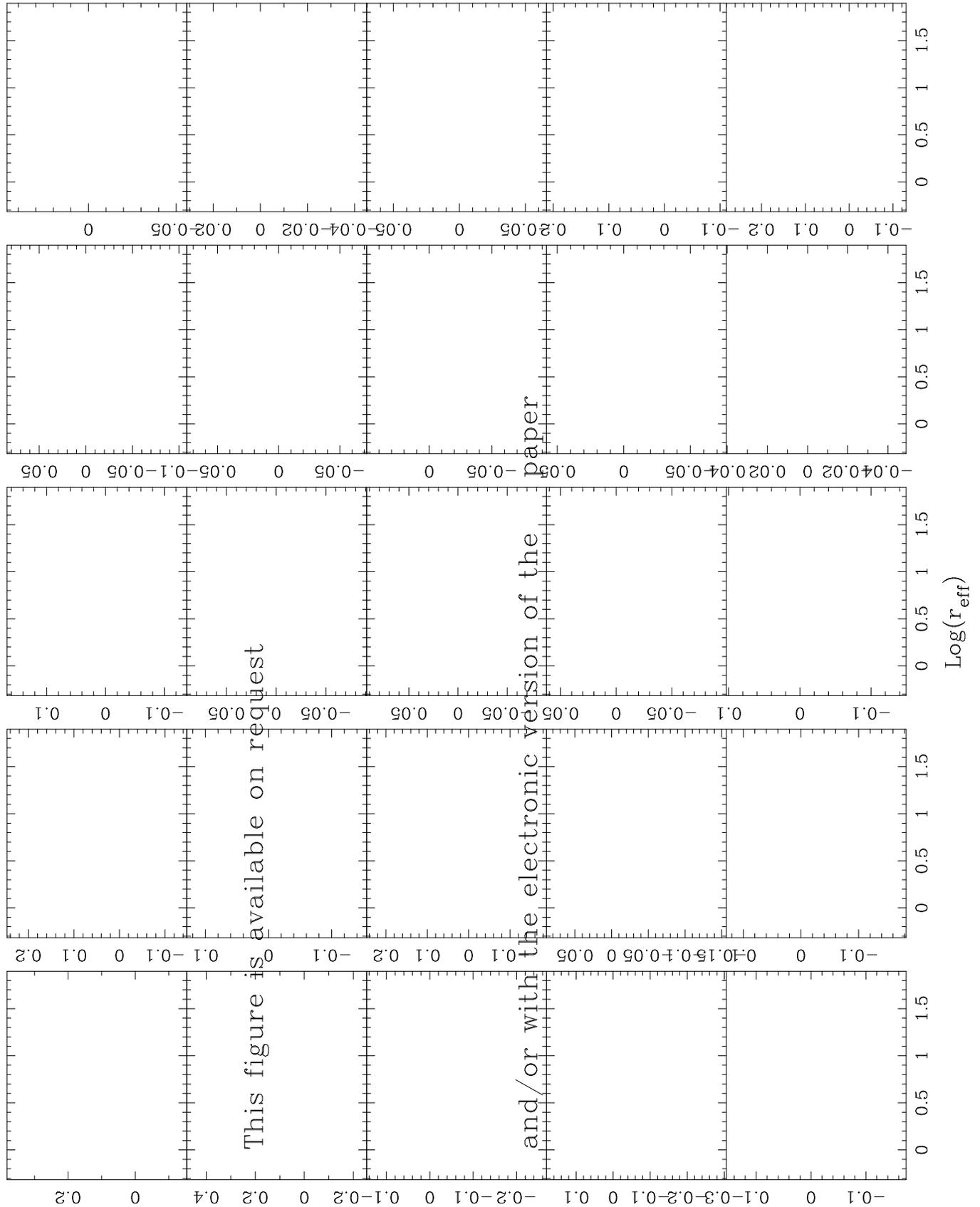}}
\caption{The relation between the gradients and the bulge 
effective radii for the 25 indices that are available for the whole
sample of bulges. The color code for the different Hubble types is the
same as in Figure \ref{sigma}.}
\label{reff}
\end{figure*}

\subsection{Building- up the gradient--$\sigma_0$ relation}

The galaxy Hubble types  are color-coded in  Figure  \ref{sigma}.  All
galaxy types  span   nearly  the  full  range  of    central  velocity
dispersions, with the  exception  of the earliest-type galaxies  (E/S0
and S0/a) whose central velocity dispersions do not reach low values.

For the E/S0  to S0/a galaxies in our  sample, the  measurement errors
are  extremely small   and therefore  give an   excellent idea  of the
intrinsic dispersion at fixed velocity dispersion.  We find that their
gradients are independent  of the bulge  velocity dispersion, whatever
index is considered.   The range of Mg$_2$  gradients in our sample is
entirely  consistent with that of   the compilation of  S0 galaxies in
Ogando et al.  (\citeyear{2005ApJ...632L..61O}).

If one only considers galaxies of types later than Sa, then the G-band
and Fe5406 indices  reveal some  variation with $\sigma_0$ as well
as Mg$_1$ and  Mg$_2$, and the slopes of  the relations  of the latter
two indices  with $\sigma_0$ are slightly stronger  than when the full
sample is considered.  Therefore, it looks  like the Mg$_1$ and Mg$_2$
trends seen  in the full  sample arise mainly from   the Sa through Sc
galaxy types.

However,  rather   than   interpreting  the  trends  between  gradient
amplitudes  and $\sigma_0$ as direct correlations,  we  find that they
are due to the combination of a gradual and differential population of
the diagrams  together with the  dynamical range of the indices (i.e.,
their intrinsic capacity to reach high values by nature).  Rather than
a smooth change in  gradient  amplitude with velocity  dispersion, one
witnesses a three-step process: (i)  At large $\sigma$, the dispersion
among gradients  is  large but  small  gradients are relatively  rare.
(ii) At  smaller $\sigma$, the  dispersion remains large, but galaxies
with very  weak gradients appear  in larger number.  (iii) Finally, at
$\sigma \la 125$  km s$^{-1}$, one finds only  bulges with either very
weak or no   radial  variation of   their  stellar population.   These
successive regimes are, of course, best seen with indices which have a
large dynamical  range of such as Mg$_1$  and Mg$_2$, but it is likely
present in all indices.

Only a few previous
studies on  gradients  included   (early-type)  galaxies with  central
velocities     below  $\sim$ 125   km     s$^{-1}$   (Gorgas  et   al. 
\citeyear{1997ApJ...481L..19G};     S\'anchez-Bl\'azquez    et     al. 
\citeyear{2006A&A...457..823S}).   Just     as  found   here,    their
low-$\sigma$ galaxies do have very  weak or null gradients.  Gorgas et
al. (\citeyear{1997ApJ...481L..19G}) find evidence for their bulges to
have  shallower  gradients than  ellipticals.   However, they  compare
systems with very different velocity dispersions.  The study of Ogando
et  al.\  (\citeyear{2005ApJ...632L..61O}) notices a decreasing number
of E and S0 galaxies  harboring steep Mg$_2$ gradients with decreasing
velocity dispersion, which is    similar  to what we find.    Finally,
Moorthy  \& Holtzman (\citeyear{2006MNRAS.371..583M}) mention that the
bulges of their sample showing insignificant  gradients in [MgFe] have
small sizes.  Histograms  of the distribution of Mg$_2$ gradients
for bulges with  $\sigma < 125$ km  s$^{-1}$  vs.\ those with  $\sigma
\geq  125$ km s$^{-1}$  are  shown in Figure~\ref{Mg2Sigmaplot}. 
Formally,    a  Kolmogorov-Smirnov  test     indicates that   the  two
distributions are different at the 99.95\% probability level.  We also
show a comparison with the elliptical galaxies of S\'anchez-Bl\'azquez
et  al. (\citeyear{2006A&A...457..823S}) which illustrates that Mg$_2$
gradients disappear  at low velocity  dispersion  in these  systems as
well.

\subsection{Other scaling relations}

Mehlert et  al.  (\citeyear{2003A&A...407..423M}) report on a stronger
correlation of absorption  line   strength gradients with  the  galaxy
velocity  dispersion {\it profiles\/}  than with  their central values. 
We indeed find a correlation, but to  a level  which is no   more
significant than  with   $\sigma_0$.   This  result holds   both  when
considering the full sample and when we restrict  our sample to S0 and
S0/a    galaxies.     This is  in    agreement  with   the  results of
S\'anchez-Bl\'azquez et al.   (\citeyear{2006A&A...457..823S})     for
elliptical galaxies. Figure \ref{reff}  displays the relations between
the  bulge gradients  and  their effective  radii.   Here again, Mg$_1$  and
Mg$_2$  are the only indices for  which a dependence  is clearly seen. 
However, the slopes  of the relations are  2\,--\,3 times smaller than
the ones  with  $\sigma_0$.  This holds  for all  indices.  This means
that any dependence  on the bulge mass  is dominated by  the effect of
the the bulge central velocity dispersion.

\section{Index--index diagrams}

We now try to understand the  nature of the  radial gradients in terms
of physical quantities such as age and/or chemical abundance
variations of the stellar population.  We  have  considered the
solar-scaled  SSP models    of  Vazdekis
(\citeyear{1999ApJ...513..224V})  and    the SSP     models      of
Thomas    et al. (\citeyear{2003MNRAS.339..897T},
\citeyear{2004MNRAS.351L..19T}) 
which  allow $\alpha$-enhancement variations.   The  use of these  two
series   of   models was   meant to   assess   the  robustness of  our
conclusions.  As a matter of fact, we find that  they indeed yield the
same broad conclusions, and we consequently choose to only present the
analysis conducted  with  the  Thomas  et  al.\ models  which  include
non-solar  [$\alpha$/Fe] values,    and thus  offer  a  more  detailed
description.

One shortcoming using  SSPs is that  when directly applied to galactic
composite  stellar    populations, the absolute    values  of  age and
metallicity   are  intrinsically  biased,    due  to  the   simplified
description as ``single'' stellar populations.  They are meant to
get close enough to the luminosity-weighted  mean quantities.  Another
problem  is the existence of  some degeneracies, in particular between
[$\alpha$/Fe]  and  age  for a fair   number   of indices.   For these
reasons, we avoid assigning absolute ages or metallicity values to our
galaxies.

As  mentioned  earlier,  galaxies  can   roughly be  divided    in two
categories: Those with significant radial gradients  and  those without.
In the following we address each  group separately  with the aim of 
possibly identify differences between them.

\subsection{Galaxies with strong gradients}

\begin{figure}[h]
\resizebox{1.0\hsize}{!}{\includegraphics[angle=-90]{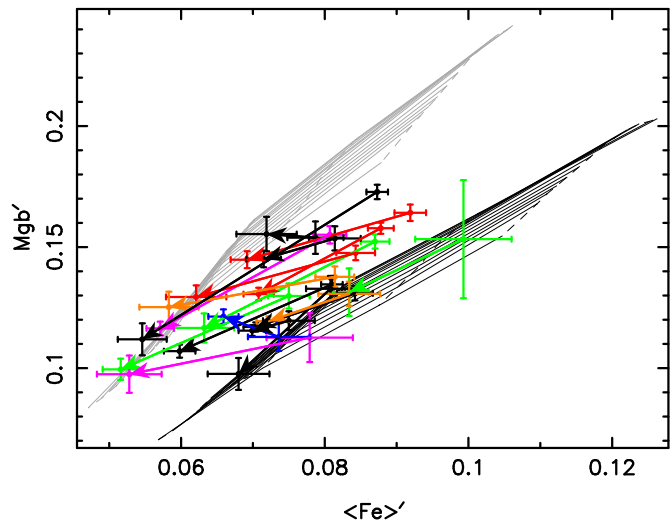}}
\caption{ Index-index diagram for galaxies harboring clear gradients. SSP Model 
grids are shown for ages running from 15Gyr to 3Gyr.  The location of
[Fe/H] $=-0.33$, 0.00, 0.35 and 0.67 are identified by dotted lines.
At the same time, they indicate the direction of an age variation.
The gray and black grids correspond to [$\alpha$/Fe]=0.3 and 0.0,
respectively. Arrows connect the central indices, integrated in a
4$''$ aperture, to the values of the indices at the effective radius
of the bulge.  The galaxy Hubble type color coding is the same as in
Figure \ref{sigma}}
\label{MgbFe}
\end{figure}

First,  we consider galaxies   for  which Mg$_2$  and  $<$Fe$>^\prime$
gradients are  strictly negative,  taking  into account   their errors
(1$\sigma$). Mg$_2$ is a  composite index, simultaneously sensitive to
metallicity, $\alpha$-elements, and age.   Therefore, this criterion does not
select   any    particular  type  of      galaxy.   The criterion   on
$<$Fe$>^\prime$  discards   3  bulges, whose   gradients,    although
negative, have large errors.   Including  them would not  help our
understanding.

\begin{figure}[h]
\resizebox{1.0\hsize}{!}{\includegraphics[angle=-90]{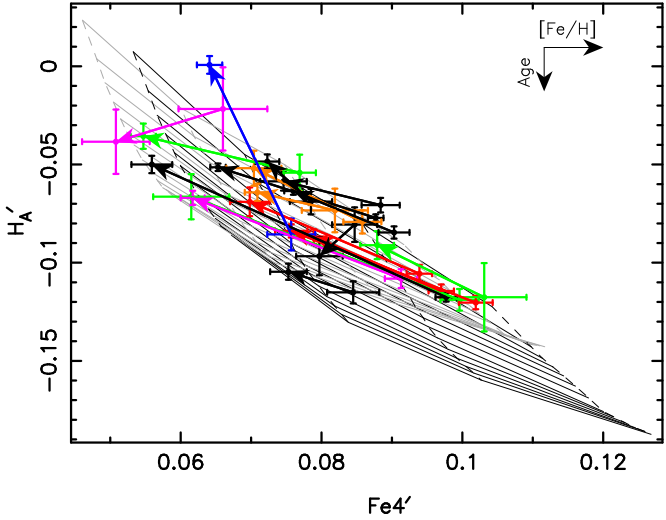}}
\caption{Index-index diagram for galaxies harboring clear Mg$_2$ gradients. The two
  grids of models for [$\alpha$/Fe]=0.0 and 0.3 are shown by black and
  light gray lines, respectively. Ages run from 15 Gyr to 3 Gyr and
  go   along  the dashed  lines  (younger  ages   correspond to higher
  H$_A^\prime$). One Gyr  separates two adjacent  age lines.  Arrows join the
  central indices,  integrated  in  a  4  arcsec aperture, with  the
  value of the indices at the bulge effective radius. The galaxy Hubble  type
color coding is the same as in Figure \ref{sigma}}
\label{HAFe4_strong}
\end{figure}

\begin{figure}[h]
\resizebox{0.9\hsize}{!}{\includegraphics{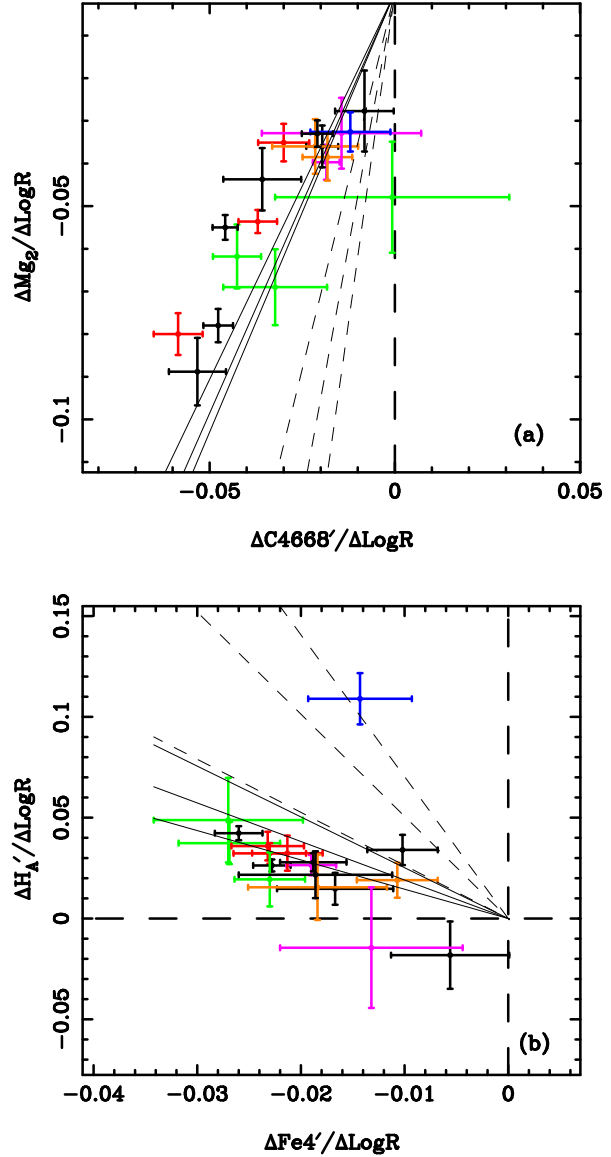}}
\caption{Gradient-gradient diagrams for galaxies harboring clear Mg$_2$
  gradients. The galaxy Hubble type color coding is the same as in
  Figure \ref{sigma}. The direction of variation of the indices with
  [Fe/H] only, or with age only, is shown by the plain and dashed
  lines, respectively.}
\label{GradGrad_strong}
\end{figure}

While the Mgb$^\prime$ vs $<$Fe$>^\prime$ diagram is highly degenerate
in age  and  [Fe/H],    it  does   offer  a good   determination    of
[$\alpha$/Fe].  In Figure  \ref{MgbFe}, for each  galaxy, we join with
an arrow  the central  indices to the  indices  measured at the  bulge
effective radius.  All galaxies lie  within a region delimited by  the
two lines at [$\alpha$/Fe]  = 0 and [$\alpha$/Fe]  =  0.3 and the  two
iso-metallicity lines  at [Fe/H]  values of  $-$0.33  and 0.67.  The
iso-age lines span the range 3- 15 Gyr.   As shown by the
direction of the arrows, most of the galaxies  show both a decrease in
metallicities and  an increase in  [$\alpha$/Fe] towards larger radii.
The variation in  [Fe/H]  is of the order   of  0.3 dex, while  it  is
smaller   for [$\alpha$/Fe]  ($\la 0.1$ dex).     A few bulges seem to
harbor pure [$\alpha$/Fe] or pure [Fe/H] gradients.

Figure \ref{HAFe4_strong} presents the variation of Fe4$^\prime$ as a
function of H$_A^\prime$ for our galaxies, superimposed on the same
grids of models as mentioned above.  Here again, the index gradients
are clearly mostly due to metallicity.  H$_A^\prime$ gradients are
small and positive. An interesting feature is that the longest arrows
(i.e., the largest gradients) are located at the bottom of the
diagram, implying a slightly older luminosity-weighted mean age.  The
relative shift is small, of the order of $\sim$\,2Gyr, but the effect
seems systematic.  Taking into account the increase in [$\alpha$/Fe]
at larger radii revealed by the previous diagnostic diagram, the
amount of possible variation due to age corresponds to a few Gyrs at
most (often of the order of 1-3~Gyr), the central parts of the bulges
being younger than their outer regions.  There are three apparent
exceptions to this general trend: UGC~11552 and UGC~11587 seem to have
stronger age gradients than the bulk of the galaxies, as they have
strong negative H$_A^\prime$ gradients.  The third galaxy, IC~5176,
harbors a strong and positive H$_A^\prime$ gradient.  The bulge of
this galaxy exhibits gradients in C4668$^\prime$ and in Mgb$^\prime$
that are among the smallest in the sample, while showing a strong
radial increase of its high-order Balmer lines towards the outer
regions of the bulge.  Its outer parts therefore seem indeed younger
than the central ones.  As a matter of fact, H$\alpha$ and [N\,{\sc
  ii}] emission are present all the way from the central to the outer
regions, which increases the uncertainties of both Balmer and Mgb
index measurements \citep[e.g.][]{1996A&A...306L..45G}. Admittedly, we
do not see any straightforward explanation for these traces of gas.
One possibility is pollution by the disk component, even though
IC~5176 does not (currently) have a bar \citep{2004AJ....127.3192C}.

Another  way to look   at the question  of the  nature of gradients is
provided by Figure \ref{GradGrad_strong}.  The SPP models, plotted in
index--index diagrams, show grids from which one can easily derive the
amount  of  variation when  only    one  of   the parameters    (age,
metallicity, or [$\alpha$/Fe])   is varying.

The advantage  of this method is  that it is immediately applicable to
any  radial    gradient--gradient  observational plot.      The slopes
associated with expected pure age and pure metallicity variations have
been derived for stellar populations of ages between  3 and 15 Gyr and
metallicities  ([Fe/H])   between $-1.35$ and    0.67.  We  considered
[$\alpha$/Fe] = 0.3,  given our  diagnosis of Figure~\ref{MgbFe}. The
results would  change in detail  but not in  conclusion if we had used
[$\alpha$/Fe] = 0.0.  The slope of the metallicity variation depends
on  the age considered and, conversely,  the  age variation depends on
the range  of  metallicity considered.  For  this  reason, we show the
maximum, mean and minimum  slopes.   Figure \ref{GradGrad_strong},
as was Figure      \ref{HAFe4_strong},   is   meant to     summarize   our
investigations. Other sets  of indices have been  considered, however,
we avoid any redundant illustration.   The choice of the two  sets
of indices,   (Mg$_2$, C4668$^\prime$)  and   (H$_A^\prime$, Fe4$^\prime$)  was
driven by four criteria:  (i) They involve well-measured  indices (ii)
the  index--index  model    grids are    regular, (iii)   they    show
significantly  different slopes for  age  and metallicity changes, and
(iv) the diagrams include sensitivities  to all three parameters (age,
[Fe/H] and   [$\alpha$/Fe]), so  that the   three can   be  considered
together.

Figure~\ref{GradGrad_strong}   demonstrates clearly   that a      pure
metallicity gradient nearly   fully   explains the amplitude   of  the
observed gradients.     Interestingly, the  smallest slopes   shown in
Figure~\ref{GradGrad_strong} fit the data the best (they correspond to
the oldest ages).  Only one galaxy, IC~5176, is more compatible with a
major radial change in age, as discussed above.  Here the line passing
through the IC~5176 location corresponds to metallicities below solar.

All galaxies  do not  fall exactly on the  models  tracing a  pure
metallicity  variation.  The residual distances to these lines are
compatible  with an age and/or [$\alpha$/Fe] variation.  As these two
parameters make  the indices move in the same direction, it is very hard 
to distinguish between them using this method, but  we already know that
likely both can be accounted for, in small proportions.

\subsection{Galaxies with weak or no gradients}

We define this category   of galaxies  as  bulges whose  gradients  in
Mg$_2$ are  compatible  with a  null value within    the errors.  Eight
galaxies in our sample fall in this  category, i.e., 25 percent of our
sample.  Their  Hubble types are mixed, but  no very early-type galaxy
is  present  (2 Sab, 2   Sb,  3 Sbc,  1  Sc).  The  three Sbc galaxies
(UGC~10043,  ESO~512-012,  NGC~1351A) have  large  error bars, both in
Mg$_2$  gradients  and    in  $\log\sigma$,  essentially   due to  the
restricted number of bins on which the gradients could be evaluated
robustly.

\begin{figure}[h!]
\resizebox{1.0\hsize}{!}{\includegraphics[angle=-90]{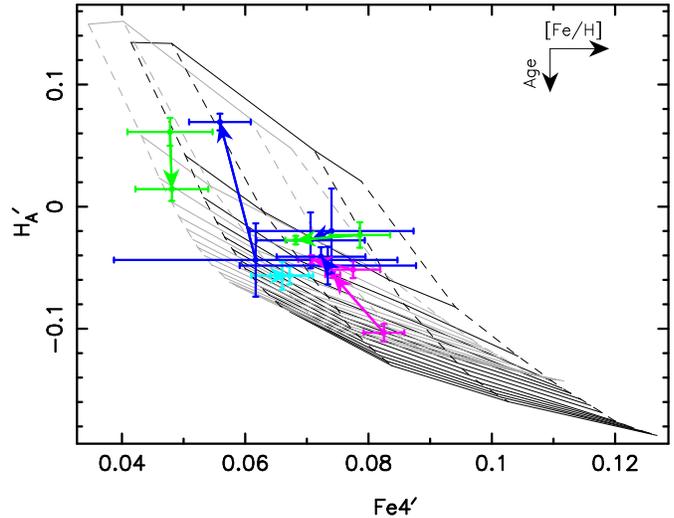}}
\caption{Index-index diagram for galaxies harboring weak or no Mg$_2$
gradient. Ages run from 1 Gyr to 15 Gyr.  The rest of the symbols and
parameters are the same as in Figures \ref{HAFe4_strong} and \ref{sigma} } 
\label{HAFe4_faint}
\end{figure}

As we did for the galaxies with strong gradients, we examine
the   relation    between   Fe4$^\prime$    and    H$_A^\prime$     in
Figure~\ref{HAFe4_faint}.   All  the bulges  with weak  gradients  are
shifted to lower  metallicities,  as compared  to the strong  gradient
ones.  

Furthermore, none of these bulges show strong Fe4$^\prime$ and
H$_A^\prime$.  This means that weak gradients are 
associated with low global chemical enrichment.  More precisely, the
indices measured at r$_{eff}$ populate a region of the
diagram which is very similar to that  of the bulges with strong
gradients (cf.\ Figure~\ref{HAFe4_strong}), but the central index
strengths do differ significantly.  This supports a scenario where
star formation proceeds from the outer to the inner parts, and where
the outer bulge regions share universal chemical properties.  The
difference between bulges arises from star formation in their central
regions.

Two galaxies    seems     to  have a    large      H$_A^\prime$ radial
variation. NGC~1351A (the  first dark blue arrow  on the  left side of
Figure~\ref{HAFe4_faint})   has amongst  the    lowest number  of   bins
available   for   the fit of its    gradients.   Despite an apparently
impressive  change  in   H$_A^\prime$, careful  consideration   of the
uncertainties  forces us to discard any  significant  gradient in this
bulge.  The case  of IC~5264  (magenta arrow) is  different.   Its
spectra have high signal-to-noise   ratios and its radial sampling  is
good.   As for IC~5176,   seen previously, we   indeed face a case  of
younger ages  in the outer parts of  the bulge.  But  again, we do not
find  any straightforward explanation for  this galaxy to be different
from the bulk of our sample.

The three remaining galaxies,  NGC~522, NGC~1886, and ESO~443-042 have
null or very weak  gradients.  They  are all  well sampled in  radius,
therefore the  absence of spectral radial  variation can be considered
real and attributable to  their small central velocity dispersion and
low global chemical enrichment.

\subsection{Comparison with elliptical galaxies}

\begin{figure}[p! h!t!]
\centerline{\includegraphics[width=8.4cm]{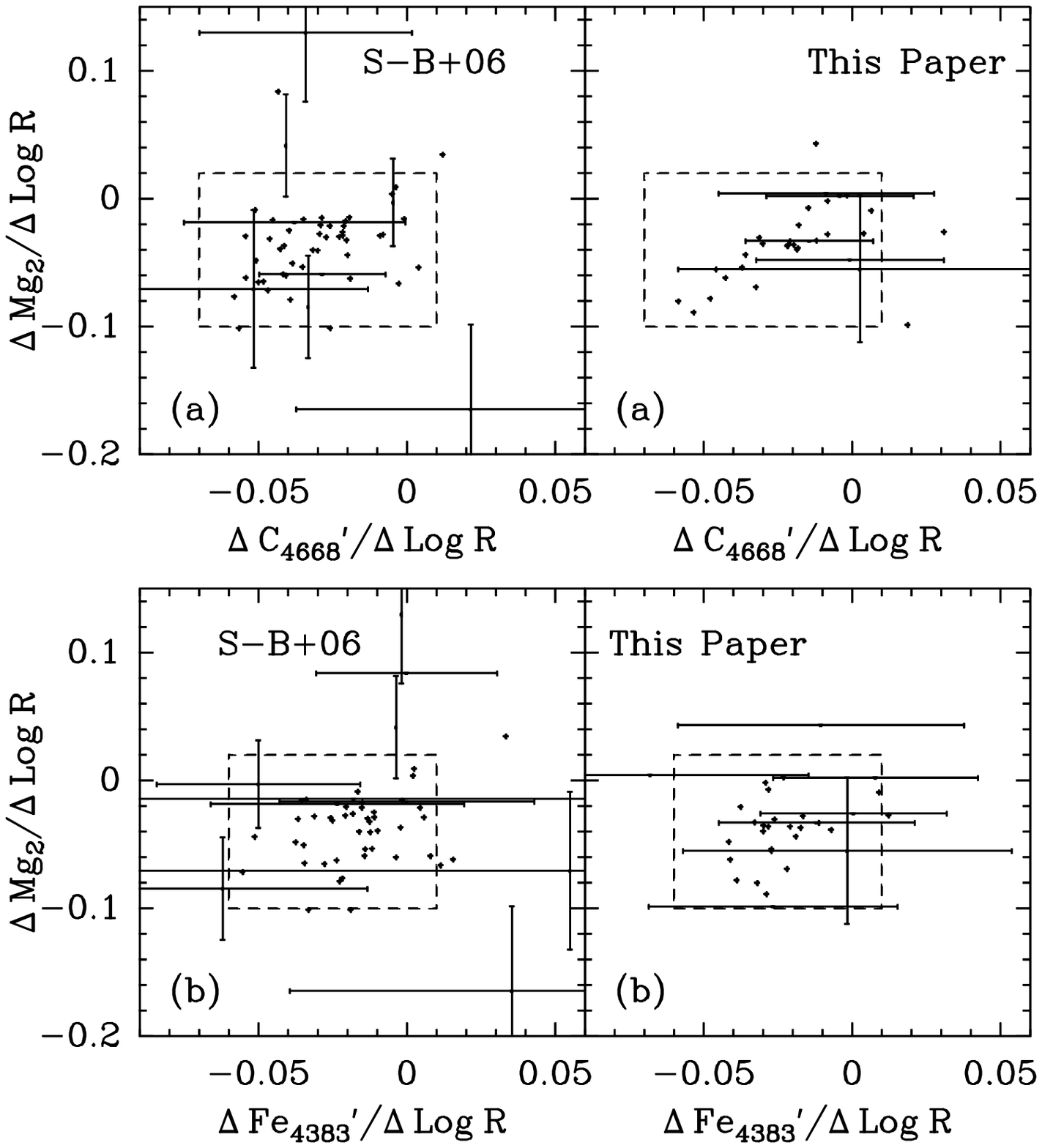}}
\centerline{\includegraphics[width=8.4cm]{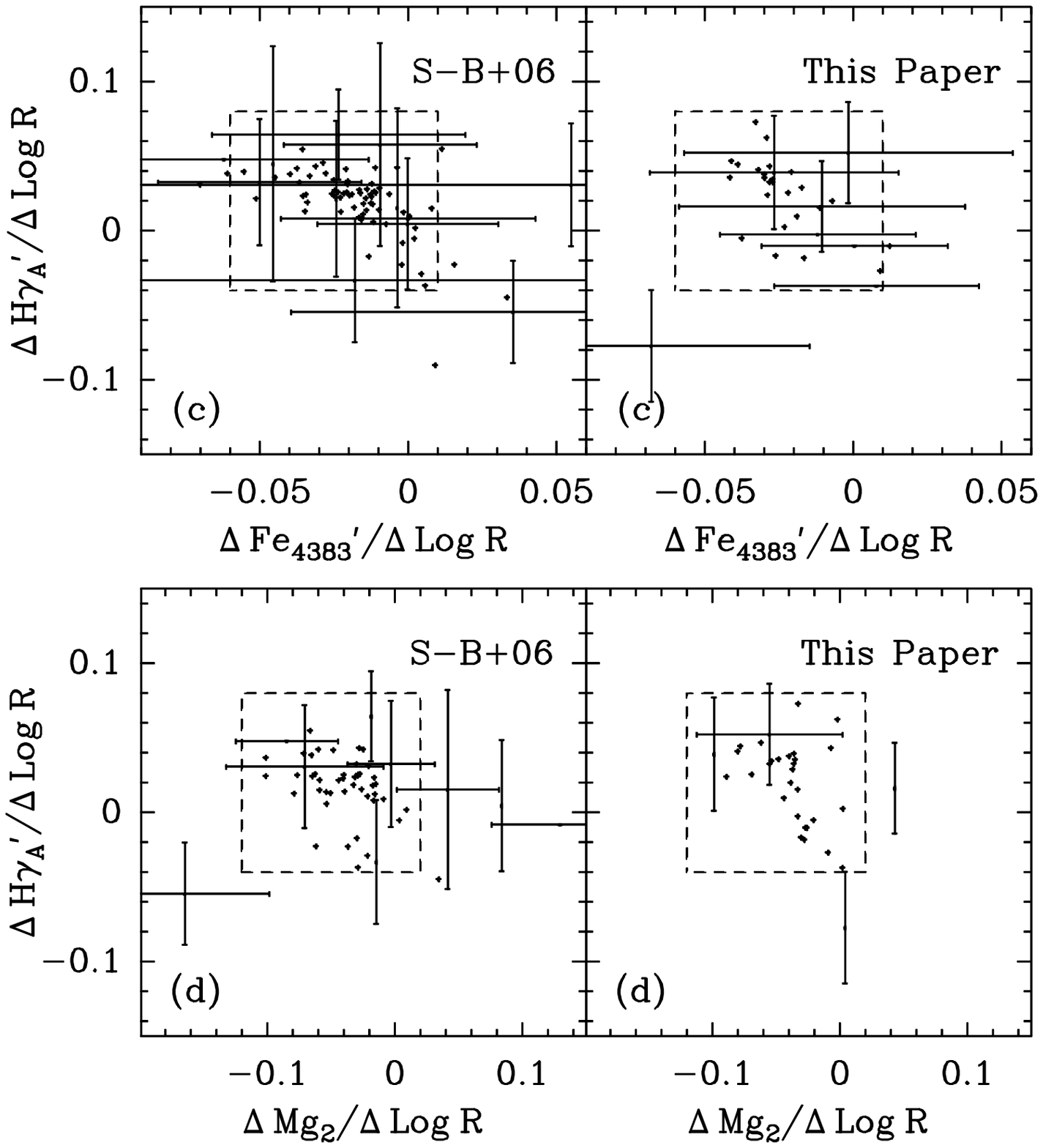}}
\caption{Comparison of  gradients of well-measured indices  for bulges  in this
paper with those for elliptical galaxies in the sample of S\'anchez-Bl\'azquez et al.\ (2006; labeled "S--B+06").  Error bars are only plotted if they exceed 0.03 dex in a given index gradient. For ease of comparison, boxes with dashed lines are drawn in a fixed position for each pair of gradient-gradient plots.}
\label{gradgradplot}
\end{figure} 

The  work  of   S\'anchez-Bl\'azquez  et  al.\  (2006)   gives  us the
opportunity to compare a set of indices  in elliptical galaxies to our
data, in a range of velocity dispersion very similar to the one of our sample
of bulges.  In particular, for our present  interest, it allows a look
at  the  relations between    the spatial distribution   of  different
indices: they  offer  clues on  the  star formation histories  in  the
galactic systems.   Figure~\ref{gradgradplot} displays this comparison
for   gradients   in  Mg$_2$,   C4668$^\prime$,    Fe4383$^\prime$ and
H$\gamma_A^\prime$.   Obviously,  the gradients in bulges  agree very
well with those  in elliptical galaxies, both in  amplitude and in the
relations between sets of   gradients.  This strengthens  further  the
similarities between the two types of spheroids.

\subsection{Quantification}

\begin{figure}
\resizebox{1.0\hsize}{!}{\includegraphics[angle=-90]{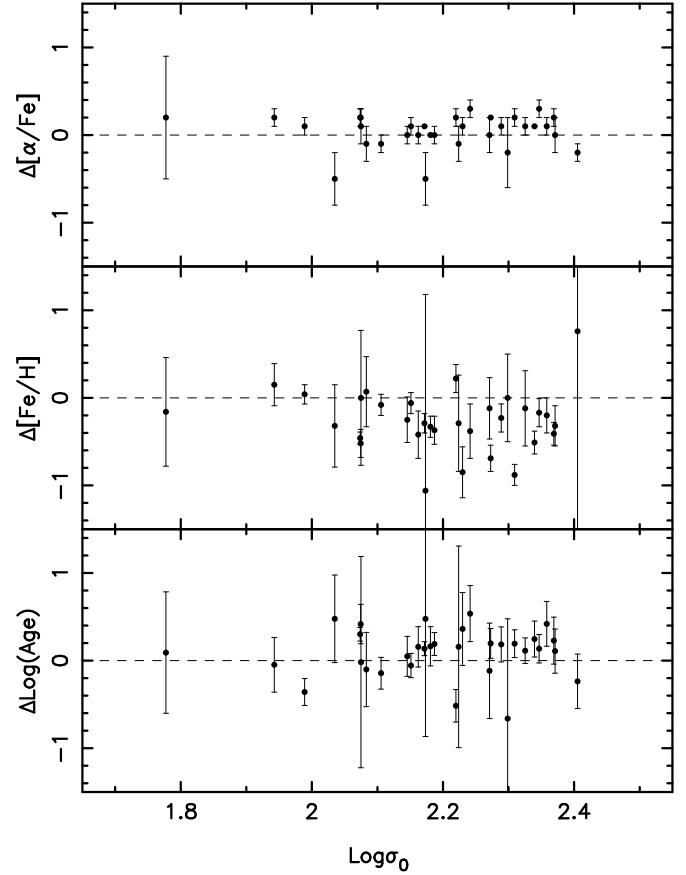}}
\caption{ The relation between $\Delta$Log(Age), $\Delta$[Fe/H] , and
  $\Delta$[$\alpha$/Fe]     with   the      bulge    central  velocity
  dispersion. $\Delta$s are measured from the bulge effective radii to
  the bulge centres}
\label{AgeFeAlpha}
\end{figure}

\begin{figure}
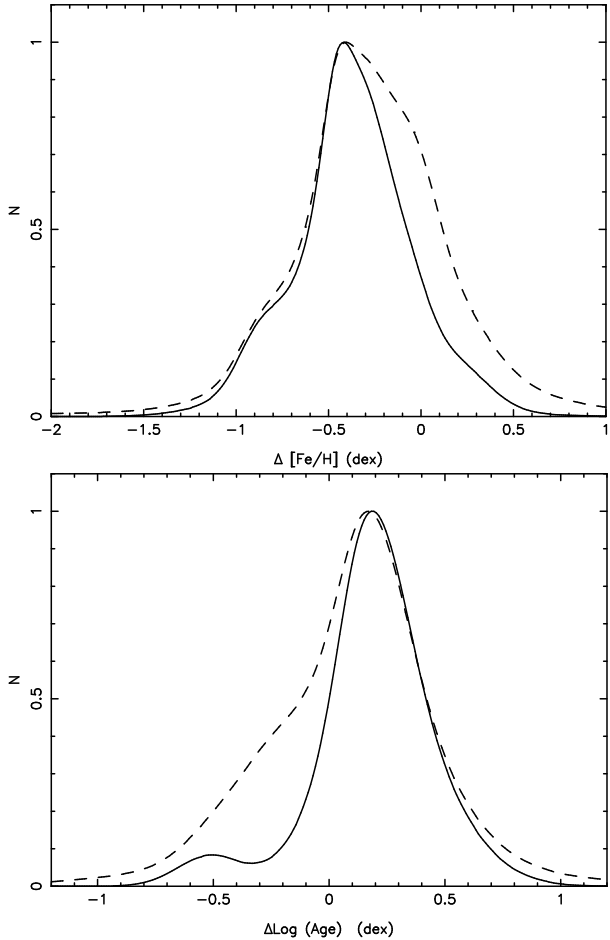

\includegraphics[angle=-90,width=8cm]{6691fig13a.ps}\hfill
\includegraphics[angle=-90,width=8cm]{6691fig13b.ps}
\caption{Generalized  histograms  of  the  distribution  of radial
changes in [Fe/H] and Log(age), calculated between the bulge effective
radii and the bulge centre, as tabulated in Table \ref{tb:AgeZ}. The
dotted lines correspond to the full sample of galaxies; the solid
lines show the histograms of the galaxies with strong gradients.}
\label{histogene}
\end{figure}

\begin{table}[h!p!t!]
\begin{tabular}{l|rcrr}
\hline
  Galaxy   &  $\Delta$[$\alpha$/Fe]  & $\Delta$ Age  & $\Delta$[Fe/H]  \\
\hline 
NGC 522     & 0.2$\pm$ 0.1  &  $-$0.6$\pm$ 4.0& 0.15$\pm$ 0.24\\
NGC 585     & 0.1$\pm$ 0.1  &  1.7$\pm$ 1.8   &$-$0.23$\pm$ 0.16 \\
NGC 678     & 0.1$\pm$ 0.1  &  6.1$\pm$ 7.6   &$-$0.85$\pm$ 0.29\\
NGC 891     &$-$0.5$\pm$ 0.3&     2.2$\pm$ 3.5&$-$1.06$\pm$ 2.24\\
NGC 973     &$-$0.2$\pm$ 0.1&   $-$6.0$\pm$ 9.5& 0.76$\pm$ 2.81 \\
NGC 1032    & 0.1$\pm$ 0.0  &   3.3$\pm$ 2.5&$-$0.51$\pm$ 0.13 \\
NGC 1184    & 0.2$\pm$ 0.1  &    2.7$\pm$ 2.9&$-$0.41$\pm$ 0.13 \\
NGC 1351A   &$-$0.2$\pm$ 0.4&   $-$4.3$\pm$14.3& 0.00$\pm$ 0.50 \\
NGC 1886    & 0.0$\pm$ 0.1  &  0.2$\pm$ 1.3&$-$0.25$\pm$ 0.26 \\
NGC 3957    & 0.1$\pm$ 0.0  &   1.0$\pm$ 0.6&$-$0.29$\pm$ 0.11 \\
NGC 5084    & 0.2$\pm$ 0.1  &   3.6$\pm$ 3.0&$-$0.88$\pm$ 0.12 \\
NGC 6010    & 0.0$\pm$ 0.0  &    2.0$\pm$ 2.6&$-$0.33$\pm$ 0.12 \\
NGC 6829    & 0.2$\pm$ 0.1  &   4.2$\pm$ 2.0&$-$0.52$\pm$ 0.16 \\
NGC 7183    & 0.1$\pm$ 0.1  & $-$0.4$\pm$ 1.0&$-$0.06$\pm$ 0.12 \\
NGC 7264    &$-$0.1$\pm$ 0.2&  0.6$\pm$ 5.4&$-$0.29$\pm$ 0.55 \\
NGC 7332    & 0.2$\pm$ 0.1&   1.1$\pm$ 1.4&$-$0.17$\pm$ 0.16 \\
NGC 7703    & 0.0$\pm$ 0.1&     1.1$\pm$ 0.8&$-$0.37$\pm$ 0.16 \\
NGC 7814    & 0.2$\pm$ 0.0&    4.0$\pm$ 3.0&$-$0.69$\pm$ 0.15 \\
IC 1711     & 0.0$\pm$ 0.2&    0.8$\pm$ 1.7&$-$0.32$\pm$ 0.23 \\
IC 1970     & 0.0$\pm$ 0.3&    2.8$\pm$ 2.8&$-$0.32$\pm$ 0.47 \\
IC 2531     &$-$0.1$\pm$ 0.2&   $-$1.0$\pm$ 4.5& 0.07$\pm$ 0.40 \\
IC 5176     & 0.2$\pm$ 0.1&  $-$4.1$\pm$ 2.5& 0.22$\pm$ 0.16 \\
IC 5264     & 0.1$\pm$ 0.1& $-$4.1$\pm$ 2.4& 0.04$\pm$ 0.11 \\
UGC 10043   & 0.0$\pm$ 0.7&    0.4$\pm$ 2.8&$-$0.16$\pm$ 0.62 \\
UGC 11587   & 0.1$\pm$ 0.1&   6.2$\pm$ 3.6&$-$0.20$\pm$ 0.20 \\
UGC11552    & 0.3$\pm$ 0.1&    6.1$\pm$ 4.9&$-$0.38$\pm$ 0.31 \\
ESO079$-$003& 0.0$\pm$ 0.2&   $-$1.1$\pm$ 6.3&$-$0.12$\pm$ 0.35 \\
ESO234$-$053& 0.1$\pm$ 0.1&    3.0$\pm$ 3.7&$-$0.12$\pm$ 0.43 \\
ESO311$-$012&$-$0.1$\pm$ 0.1 & $-$1.1$\pm$ 1.5&$-$0.08$\pm$ 0.12 \\
ESO443$-$042& 0.0$\pm$ 0.1&    0.7$\pm$ 0.9&$-$0.42$\pm$ 0.27 \\
ESO512$-$012& 0.1$\pm$ 0.2&    $-$0.1$\pm$ 7.2& 0.00$\pm$ 0.77 \\
\hline
\end{tabular}
\caption{Variation in [$\alpha$/Fe], age and [Fe/H] between the bulge
  bulge effective radii and the central regions. Negative values indicate
an increase of the considered quantity. }
\label{tb:AgeZ}
\end{table}

We now try to  quantify the [$\alpha$/Fe],  age and  [Fe/H] variations
between the bulge  effective radii and their central  parts .  We  use
the models of Thomas et al.  (\citeyear{2003MNRAS.339..897T},
\citeyear{2004MNRAS.351L..19T}).  First, we derive [$\alpha/{\rm Fe}$]
from  the   Mgb$^\prime$-$<$Fe$>^\prime$ diagram.    Subsequently  and
independently, we calculate the age and  metallicity, as inferred from
the  H$_A^\prime $--Fe4$^\prime$ plane, at  r$_{eff}$ and in the bulge
central parts.  The ages  and metallicities are derived at
the computed [$\alpha/{\rm Fe}$] values. For those, we have considered
initial mean ages of 10 and 5 Gyr, and  checked that the dependence of
the  final age and  metallicity variation on  this a priori choice was
negligible.

The results are listed in  Table~\ref{tb:AgeZ}.  As noticed previously
in a qualitative  way, the outer parts of  the bulges have (with a few
exceptions) higher   [$\alpha/{\rm    Fe}$], older ages     and  lower
metallicities than the  inner parts.  Only  two galaxies, IC~5264  and
IC~5176, have significant positive  age gradients.   Note that due  to
the  uneven   spacing between   the model   lines  in  the H$_A^\prime
$--Fe4$^\prime$ diagram, the absolute errors attached to variations in
age are  larger when  measured  for the  oldest populations (for which
model  lines   are  close  together)   than for   younger populations. 
Similarly, the variations themselves are maximized.  Therefore, we are
effectively measuring upper limits of  $\Delta$age.  This effect  does
not exist  for [Fe/H]  for   which the grid    lines are more   evenly
separated from one  another.  The uncertainties in  $\Delta$[Fe/H] are
mainly linked  to the errors  in $\Delta$[$\alpha/{\rm  Fe}$] (besides
the observational ones).  This is particularly true for galaxies which
seem  to  harbor a  positive   metallicity gradient.  Galaxies  having
insignificant   age   gradients  also  have   negligible   metallicity
gradients.   Figure~\ref{AgeFeAlpha}    illustrates  the   results  of
Table~\ref{tb:AgeZ} and presents the  relations between the variations
in [$\alpha/{\rm Fe}$], age and [Fe/H] with the bulge central velocity
dispersion.  The description of the Mg$_2-\sigma_0$ relation in Sect.\ 
5.2  is  applicable  here as well.    This  is particularly   true for
$\Delta$[Fe/H] which has the largest dispersion: I.e., there isn't any
significant correlation  between   metallicity gradient  and  velocity
dispersion, strictly  speaking.   Instead one sees  a  decrease in the
range of possible gradient values at lower $\sigma$.

In  order  to quantify   the qualitative statement  made  earlier that
bulges have comparable properties at their effective radii, while they
differ  in  their  central regions,  we  measured their  mean ages and
metallicities.  At identical ages,  the weak  gradient bulges have  an
error weighted mean [Fe/H]($r_{\rm eff}$)= 0.11 dex, with a dispersion
of  0.14  dex.  Strong   gradient  bulges have  a mean  [Fe/H]($r_{\rm
  eff}$)=0.05    dex with a  dispersion of    0.19 dex.   This must be
compared to   a   mean central metallicity   of  [Fe/H]   = 0.13   dex
(dispersion of 0.19  dex) and 0.45 dex, (dispersion  of 0.14 dex), for
the weak and strong  gradient  bulges, respectively.  Note that  these
values are derived from  SSP models.  As  such, they are not definite. 
The comparison  they  allow between galaxies is   however robust.  The
galactocentric distance of one effective radius  is chosen because all
our galaxies have spectra of adequate quality there.  However, it does
not yet sample the most outer bulge regions, where we think there is a
lot to learn on the very early stages of bulge formation.

Figure~\ref{histogene}  shows  the generalized  histograms   of 
$\Delta$Log(age) and $\Delta$[Fe/H].   We  choose here   a logarithmic
representation of the age for   internal homogeneity (in the sense that
index variations scale approximately linearly with Log (age))  and
for a more direct comparison with the metallicity scale. Both the full
sample, in dashed line, and the  strong-gradient galaxy sub-sample, in
solid line,  are shown. By nature of the use of generalized histograms,
the errors  are  taken  into account.  

Both the age and the  metallicity gradient distributions are very well
peaked,  at  $\sim$0.15 dex in   Log(age)   (1.5~Gyr  in  age) with   a
dispersion $\sigma=0.2$ dex   (in Log(age)) (1.3  Gyr  in age),  and at
$-0.4$  dex with  a  dispersion  $\sigma=0.3$ dex  for   [Fe/H].  They
correspond to   peak  values  of  the  gradients  of   $\sim$0.07 and
$\sim-$0.2 for the Log(age) and [Fe/H], respectively.

Similarly, the mean gradients
for the age and  [Fe/H] are 0.06 and  $-$0.16, respectively.  
In other words, metallicity gradients  are two to three  times larger than
those in age (in log  scale).  These values are very  close to those
derived for elliptical galaxies with a  comparable range of velocity
dispersions by S\'anchez-Bl\'azquez et al.
(\citeyear{2006A&A...457..823S},   see 
summary of other analyses therein): 0.082 and $-$0.206.  The different
grids  of   models and relative   distribution in  velocity dispersion
within each sample suffice in explaining the small differences between
their and our analyses.  Although the distribution of gradient amplitude is
more meaningful than their mean values in studying the processes at play in
building spheroids, the similarity between mean values of gradients in
ellipticals and bulges suggests that they must share  a fair amount
of common formation history.

\section{Conclusions}

We   have presented the  analysis   of  radial  gradients of   stellar
absorption   line strengths in a  sample   of 32 edge-on spiral galaxy
bulges. The sample galaxies span nearly the full Hubble sequence (from
S0 to Sc types), and have a  large range of dynamical properties, with
their  bulge central velocity  dispersions  ranging from $\sim$\,60 to
300 km~s$^{-1}$.  Our main conclusions are the following:

\noindent $\bullet$ Most  bulges do present  radial stellar population
gradients.   The  outer   parts of   bulges harbor  weaker metallic
absorption lines than the  inner  regions.  The distribution of  these
gradients among bulges are generally well peaked.  They also display a
real intrinsic dispersion, implying the presence of  a variety of star
formation histories within a common framework.

\noindent $\bullet$ In a number of cases, the gradients in Mg$_1$
and Mg$_2$ do not follow those in Mgb.  We suggest that this is due to
the  inclusion  of carbon,  traced   by  the   C4668$^\prime$ index,  in  their
bands. This explains a   number  of apparent  previous   discrepancies
between works  dealing either with  Mg$_2$  or Mgb, in particular  for
elliptical galaxies.

\noindent $\bullet$ We argue that the existence  of a {\it correlation\/}
between gradients and central velocity dispersion  among bulges is not
an appropriate terminology.  Instead, one sees  that bulges with large
velocity dispersion can exhibit strong  gradients (they can also  have
negligible ones), while this probability diminishes at lower $\sigma$.
Below 125 km~s$^{-1}$,  we do not find  any bulge with significant
radial spectral  gradient.   The  same kind of   dual  regime with the
velocity   dispersion  was observed in    elliptical galaxies.  This
gradual  build-up  of the  index vs.\  $\sigma$   relation can only be
clearly observed for indices with large dynamical ranges.

\noindent $\bullet$ The Hubble type of the  parent 
galaxy appears  to be a   secondary    parameter, earlier  types   having
statistically larger  velocity  dispersions   than later   ones.   The
strength of the  gradients depends only weakly  on the bulge effective
radius.  The difference in sensitivity to  the effective radius on one
hand, and to the velocity dispersion on  the other hand, suggests that
the depth of the gravitational  potential in which bulges are  embedded
is the main property tracing the spatial distribution of their stellar
population.

\noindent $\bullet$ Strong  gradients are  found  in bulges with  the  most
metal-rich central regions, while  bulges with strong {\it and weak\/}
gradients have comparable  properties at  the bulge effective  radius.
This indicates  that the external   regions of  the bulges, which  are
typically the oldest  and most metal-poor and  therefore the  first to
form,  share some universal properties.  This  also argues in favor of
an outside-in scenario for the formation and evolution of bulges.

\noindent $\bullet$ The analysis using SSP models  indicates that
radial variations in luminosity-weighted mean metallicity are twice to
three time as  large (in logarithmic scale) as  the variations in age.
While [Fe/H]  at  the bulge  effective  radii  are on  average  0.4 dex
($\sigma$ = 0.3 dex) lower  than in the  bulge central regions, the age
difference is of the order of 1.5 Gyr  ($\sigma$ = 1.3 Gyr), the inner
regions being   younger.    We have  found  only    two galaxies  with
convincing  signs of  an `inverted'   age gradient.   The changes   in
[$\alpha$/Fe] are small  (of the order   0.1 dex) and  rather constant
among bulges.

As to the fundamental question of the influence of the disk on the
bulge evolution, we make the following comments: In terms of stellar
population properties, we do not find any obvious differences between
ellipticals and bulges.  On the contrary, it seems that our
observations strengthen their resemblance.  One requirement to
properly compare the two families of spheroids is to match systems of
similar velocity dispersion. The imprint of the disk influence is
generally seen via the presence of a bar in spiral galaxies.
\cite{2005MNRAS.364L..18B} determined that a cycle of bar formation
and dissolution takes about  2 Gyr for  Sb-Sc galaxies with masses and
radii comparable to that  of the Milky Way.   Gas accretion allows the
bar to reappear \citep{2002A&A...392...83B}.  In terms of time scales,
a scenario  where  bulges would be   formed through secular  evolution
could thus be conceivable for   late-type galaxies.  However, for  the
time  being, our study introduces  two severe constraints to this type
of scenario: (i)  The  action of the   bar must preserve the  observed
gradients  and star  formation time  scales (as  measured by
[$\alpha$/Fe],  whose  measured values  are  significantly larger than
those of disk stars); (ii)  The age gradients produced  by the bar
cycles  (formation to dissolution) must  be consistent with those
measured (peaked at  $\sim$1.5 Gyr).  For example, if  2  to 3 bar
cycles are necessary  to build the bulge of  an early-type galaxy from
disk    material, the bulge assembly    timescale  is larger than  the
one implied by the observed gradients.

We  note  that these remarks also  hold for models   where  bulges grow by
accretion of satellites, combined with  the accretion of disk material
(e.g., \cite{2006A&A...457...91E}).  Nevertheless, bars are ubiquitous
among spirals, and their presence and action could explain part of the
dispersion  seen  among  the  bulge    properties  at fixed   velocity
dispersion, in particular by  wiping out strong  gradients \citep[see,
e.g.][]{1994ApJ...430L.105F}.   Contrary   to  what   was   found   by
\cite{2006MNRAS.371..583M}, we do not  find  that younger ages in  the
centres of  bulges are restricted to   barred and/or boxy/peanuts shape
galaxies.  In  fact, we do  not find any  difference between the stellar
properties of  barred and non-barred galaxies  in our sample,  as seen
along their bulge minor axis.   

\cite{2004MNRAS.347..740K}   has  proposed,    as a   result    of her
simulations of elliptical galaxies, that galaxies with strong
gradients originate from small and rapid mergers mimicking an initial
(monolithic) collapse, whereas galaxies with weak gradients would
reveal a sequence of more significant mergers (i.e., with
small mass ratios).  In a scenario where bulges and ellipticals form
in a similar way, i.e., disks settling after the bulk of the stellar
population in bulges is in place, this hypothesis would also fit part
of the observations presented here. However, bulges with low velocity
dispersions would then require a different formation mechanism.

\begin{acknowledgements}
This work was supported by the Spanish research project AYA 2003-01840
\end{acknowledgements}

\bibliographystyle{aa}
\bibliography{astro6691}

\end{document}